\documentclass[a4paper]{article}

\usepackage{graphicx}      
\usepackage{amsmath}       
\usepackage{amsfonts}
\usepackage{amssymb}
\usepackage{cite}
\usepackage{booktabs}      
\usepackage{caption}
\usepackage{subcaption}
\usepackage{threeparttable}
\usepackage{upgreek}
\usepackage{soul}
\usepackage{siunitx}
\usepackage[utf8]{inputenc}
\usepackage[T1]{fontenc}
\usepackage{newtxtext,newtxmath} 
\usepackage[left=3.75cm,right=2.5cm,top=3.75cm,bottom=3.75cm]{geometry}
\usepackage{setspace}
\usepackage{hyperref}
\hypersetup{colorlinks,linkcolor={red},citecolor={blue},urlcolor={red}}
\usepackage{mwe}           
\usepackage{float}         
\usepackage{authblk}
\usepackage{titlesec}
\usepackage{tabularx}
\usepackage{graphicx}
\usepackage{array}
\usepackage{makecell}
\newcolumntype{C}{>{\centering\arraybackslash}X}
\usepackage{bm}

\titleformat{\section}
{\normalfont\normalsize\bfseries}
{\thesection}{1em}{}

\titleformat{\subsection}
{\normalfont\normalsize\bfseries}
{\thesubsection}{1em}{}

\titleformat{\subsubsection}
{\normalfont\normalsize\bfseries}
{\thesubsubsection}{1em}{}

\titleformat{\paragraph}
{\normalfont\normalsize\bfseries}
{\theparagraph}{1em}{}
\begin{document}
	\date{}
	\singlespacing
	\graphicspath{{Figures_Ni_Doped/}}
	\title{Influence of Ni Doping on the Structural, Morphological, Optical, and Electrical Properties of Nanocrystalline $\mathrm{Cd_{1-x}Mn_xS}$ Thin Films}
	
	\author[1]{Himanshu Sharma Pathok\thanks{Corresponding author: rs\_hspathok@dibru.ac.in}}
	\author[2]{Padma Pani Shahu}
	\author[3]{Himanshu Kalita}
	\author[4]{Prasanta Kumar Saikia}
	
	\affil[1,2,3,4]{Thin Film Laboratory, Department of Physics, Dibrugarh University, Dibrugarh, Assam-786004, India}
	
	\maketitle
	
	\doublespacing
	\begin{abstract}
		Ni-doped $\mathrm{Cd_{1-x}}\mathrm{Mn_x}\mathrm{S}$ (x = 0.4) thin films were prepared via a cost-effective chemical bath deposition (CBD) method to investigate their suitability for optoelectronic applications. Incorporation of a secondary transition metal such as Ni is expected to influence lattice strain, defect density, and electronic structure through ionic size effects and sp–d exchange interactions, thereby providing an additional degree of freedom for tuning the properties of $\mathrm{Cd_{1-x}}\mathrm{Mn_x}\mathrm{S}$ -based ternary systems. X-ray diffraction (XRD) analysis confirmed the cubic zinc blende structure of the $\mathrm{Cd_{1-x}}\mathrm{Mn_x}\mathrm{S}$ crystal, which was further corroborated by high-resolution transmission electron microscopy (HRTEM). The crystallinity and crystallite size were found to increase with increasing Ni concentration. The size of the spherical grains also increased with increasing Ni doping concentration. Field emission scanning electron microscopy (FESEM) analysis revealed uniform, dense, and crack-free films with grain size increasing as a function of Ni content, and the FESEM cross-sectional images indicated a nearly constant thickness in the range of ~ 181.2–189.1 nm. The films exhibited high optical transmittance (75–90\%) in the visible and near-infrared (NIR) regions. The optical band gap decreases from ~2.72 to 2.62 eV as the Ni concentration increases from 1\% to 4\%. Current–voltage (I–V) measurements revealed enhanced electrical conductivity, which further increased under illumination, confirming the photoconducting nature of the films. Overall, these results demonstrate that Ni doping effectively tailors the properties of $\mathrm{Cd_{1-x}Mn_xS}$ thin films, suggesting their potential for optoelectronic applications and indicating their possible suitability as window layer materials for thin-film solar cell.
	\end{abstract}
\textbf{keywords}: Chemical Bath Deposition; Cadmium manganese sulfide; Thin film; X-ray diffraction; High resolution transmission electron microscopy
\section{Introduction}
Ternary thin films have attracted significant attention due to their tunable energy band gap, adjustable effective mass, and the variation of optical and structural properties with changes in material composition. Owing to these versatile characteristics, ternary thin films are widely employed in applications such as sensing devices \cite{sivaperuman2024binary} as well as thin-film solar cells (TFSCs) \cite{gayathri2025fabrication}. In recent years, transition metal chalcogenide thin films, in particular, have gained prominence in optoelectronic applications due to their relatively high band gaps and favourable electronic properties \cite{khan2021introduction}. Cadmium sulfide (CdS) is one of the most widely studied II–VI semiconductor materials and is extensively used in optoelectronic applications such as photodetectors \cite{li2010single}, gas sensors \cite{yadava2010sensing}, window layers in polycrystalline thin-film solar cells (TFSCs), etc \cite{coutts1982high}. These applications are mainly attributed to its high optical transmittance, large absorption coefficient, and direct band gap of about 2.42 eV, which make CdS suitable as a window layer material for TFSCs \cite{saikia2011synthesis, bora2021effect}. However, this band gap leads to absorption losses in the blue and ultraviolet regions of the solar spectrum, thereby reducing the short-circuit current density (Jsc) of TFSC devices \cite{borah2025effect} . In addition, the toxicity of cadmium raises serious environmental and health concerns \cite{zyoud2018recycled, hannachi2016growth}.
Manganese sulfide (MnS) thin films have been proposed as an alternative window layer material to overcome these limitations. MnS exhibits high optical transmittance, a suitable absorption coefficient, and a wide direct band gap in the range of 3.1–3.7 eV, which makes it attractive for optoelectronic applications \cite{shokr2024optical,ulutas2013gamma}. Among ternary II–VI semiconductor systems, Mn-based thin films are of particular interest because Mn ions, owing to their comparable ionic radius and half-filled d-shell electronic configuration, can substitute Cd lattice sites in Cd-based materials and thereby induce controlled modifications in structural, optical, and electrical properties. Consequently, both CdS and MnS have been extensively investigated for optoelectronic device applications.\\The incorporation of Mn into CdS leads to the formation of ternary Cadmium manganese sulfide ($\mathrm{Cd_{1-x}}\mathrm{Mn_x}\mathrm{S}$) thin films, which exhibit enhanced and tunable properties compared to the corresponding binary compounds \cite{Jawad}. By varying the Mn concentration, key parameters such as band gap energy, lattice constant, and charge carrier characteristics can be systematically tailored, enabling optimization for different device applications \cite{chuu1997growth}. As a result, the deposition of $\mathrm{Cd_{1-x}}\mathrm{Mn_x}\mathrm{S}$ thin films on suitable substrates has received significant attention, as thin-film growth provides improved control over material properties at the nanoscale and supports diverse optoelectronic and photovoltaic applications.\\
Metal chalcogenide thin films can be deposited using various techniques, such as sol-gel dip coating \cite{Goktas1, Goktas2}, Chemical bath deposition (CBD) \cite{Lihare}, Spray pyrolysis \cite{Yuksel}, RF Sputtering \cite{Das}, spin coating \cite{Rahman}, Chemical vapor deposition \cite{Uda}, Successive Ionic Layer Adsorption and Reaction (SILAR) \cite{Mukherjee}, etc. Among these methods, CBD stands out as a cost-effective, simple, and scalable approach \cite{mugle2016short, Lihare}. It does not require sophisticated instrumentation or vacuum systems, making it highly accessible and well-suited for large-scale production \cite{Arandhara1}. In CBD, the properties of the deposited films can be precisely controlled by varying parameters such as precursor concentration \cite{kakhaki2022influence}, deposition temperature \cite{hasan2023effects}, solution pH \cite{murthy2022effect}, and deposition time \cite{adhyapak2024chemically}. Furthermore, its low-temperature processing makes CBD especially advantageous for large-area coatings in industrial applications.
\\Extensive research has focused on tuning the properties of semiconductors through rare-earth and transition metal doping, which can significantly alter their optical and electrical characteristics and thereby enhance their functional performance and application potential \cite{veeramanikandasamy2016effect, Susha}. Moreover, the type of semiconductor, whether n-type or p-type, can be modified through appropriate doping techniques. The doping of transition metals into ternary thin films has not been widely investigated. Among transition metals, $\mathrm{Ni^{2+}}$ is of particular interest due to its ferromagnetic nature \cite{murugesan2017structural}, which may introduce unique magnetic and electronic properties when incorporated into semiconductor materials \cite{horoz2017structural}. N. Naveenkumar et al.\cite{naveenkumar2025improved} reported that Ni doping enhances charge carrier mobility, accelerates charge transport, and improves the electrical conductivity of semiconductor thin films. Similarly, M. Younus Ali et al. \cite{ali2020effect} demonstrated that transition-metal doping induces metallic characteristics, noting that Ni, owing to its high electrical and thermal conductivity and moderate electronegativity, exhibits a strong affinity for bonding with host lattice elements. I. S. Yahin et al. \cite{yahia2019chemically} further showed that Ni ions can readily penetrate the CdS lattice due to their smaller ionic radius, leading to significant improvements in the electrical performance of the thin films. S. Chandramohan et al. \cite{Chandramohan} and R. Bairy et al. \cite{Bairy} investigated the effect of Ni incorporation on CdS thin films and demonstrated that Ni doping significantly influences their structural, optical, and morphological characteristics. Both studies reported a decrease in the optical band gap energy with increasing Ni concentration, accompanied by an increase in grain size, which was attributed to enhanced grain agglomeration at higher Ni doping levels. These findings suggest that Ni is a promising dopant capable of effectively tailoring the structural, morphological, optical and electrical properties of semiconductor materials. Despite numerous studies on Ni-doped CdS thin films, corresponding investigations on Ni-doped $\mathrm{Cd_{1-x}Mn_xS}$ thin films have not been reported in the literature. Therefore, the present study provides the first systematic investigation of the influence of Ni incorporation on $\mathrm{Cd_{1-x}Mn_xS}$ thin films prepared by the cost-effective chemical bath deposition technique.\\ In our previous study, $\mathrm{Cd_{1-x}Mn_xS}$ thin films with x = 0.4 exhibited promising optical and electrical properties, indicating their suitability for optoelectronic applications \cite{pathok2025effect}. Therefore, the present study focuses on investigating the effect of Ni doping on $\mathrm{Cd_{1-x}Mn_xS}$ (x=0.4) thin films. The main objective is to examine how varying Ni concentration influences the physical properties of the films and to assess their potential for enhanced optoelectronic performance. 
\section{Preparation technique}
\subsection{Preparation of substrate}
Ni doped $\mathrm{Cd_{1-x}}\mathrm{Mn_x}\mathrm{S}$ thin films were deposited onto microscopic soda lime glass substrates. The polycrystalline thin films are typically deposited on amorphous glass slides or quartz \cite{liu2023plasma}. In this study, microscopic glass slides (Riviera) with dimensions of ($25\times25\times1$) $\mathrm{mm^3}$ were used as substrates. The glass slides were first cleaned with a neutral liquid laboratory detergent (Labolene) and then rinsed with deionised water. After cleaning, the slides were dried in an oven at approximately 150°C. Following a 4-hour drying period, the slides were immersed in a 10\% dilute nitric acid solution. Acid treatment is crucial for removing both organic and inorganic contaminants. After one hour of acid treatment, the slides were again washed with detergent and subsequently placed in an ultrasonic bath with deionised water. The ultrasonication process helps to remove fine particles adhering to the glass surface through the action of high-energy microbubbles generated in deionised water by high-frequency sound waves. Finally, the slides were dried using a hot air dryer.
\subsection{Preparation of bath solution}
\begin{figure}[ht]
	\centering
	\includegraphics[width=0.8\textwidth,height=0.3\textheight]{"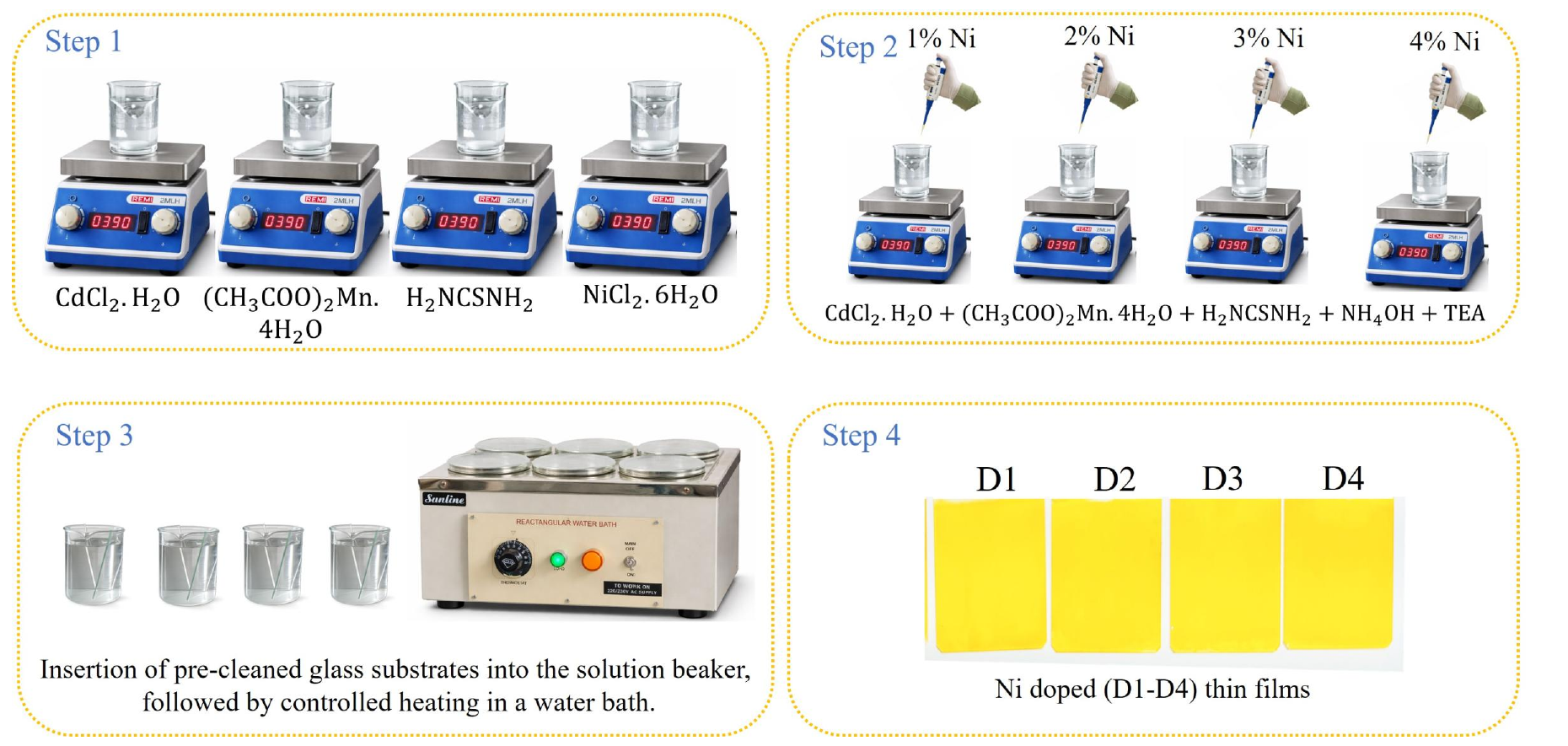"}
	\caption{Schematic diagram for preparation process of Ni doped $\mathrm{Cd_{1-x}Mn_xS}$ thin films}
	\label{fig:Fig_1}
\end{figure}

Cadmium chloride monohydrate [$\mathrm{CdCl_2 \cdot H_2O}$], manganese acetate tetrahydrate [$\mathrm{Mn(CH_3COO)_2\cdot 4H_2O}$], and thiourea [$\mathrm{H_2NCSNH_2}$] were used as the cadmium, manganese, and sulfur precursors, respectively. For Ni doping, nickel chloride hexahydrate [$\mathrm{NiCl_2 \cdot 6H_2O}$] was employed. Ammonia solution (30\%) [$\mathrm{NH_4OH}$] and triethanolamine (TEA) [$\mathrm{C_6H_{15}NO_3}$] were used as complexing agents to control the reaction mechanism and suppress the formation of unwanted precipitates.\\All precursor solutions were prepared with a concentration of 0.2~$\mathrm{mol\,L^{-1}}$. Initially, 6~mL of $\mathrm{CdCl_2\cdot H_2O}$ was mixed with 4~mL of $\mathrm{Mn(CH_3COO)_2\cdot 4H_2O}$. The required amounts of $\mathrm{NH_4OH}$ and TEA were then added until a clear solution was obtained, indicating proper complex formation. Subsequently, 20~mL of $\mathrm{H_2NCSNH_2}$ was added to the solution. In total, four such solutions were prepared.\\Finally, Ni was doped into the four beakers by adding 0.1~mL, 0.2~mL, 0.3~mL, and 0.4~mL of 0.2~$\mathrm{mol\,L^{-1}}$ $\mathrm{NiCl_2\cdot 6H_2O}$ solution, corresponding to 1\%, 2\%, 3\%, and 4\% volume doping relative to the total volume of the $\mathrm{CdCl_2 \cdot H_2O}$ and $\mathrm{Mn(CH_3COO)_2\cdot 4H_2O}$ precursor solutions, as shown in Table \ref{tab:1}. The pH of all solutions was maintained at $(10.00 \pm 0.20)$. The deposition technique employed for the fabrication of the films is illustrated in Figure \ref{fig:Fig_1}. To assess the reproducibility of the deposition process, three independent samples were prepared for each Ni doping concentration under identical deposition conditions. The samples exhibited consistent structural, optical, electrical and morphological characteristics and a representative sample from each composition was selected for detailed characterization.
\subsection{Chemical reactions}
Ni-doped $\mathrm{Cd_{1-x}}\mathrm{Mn_x}\mathrm{S}$ thin films were deposited via the chemical bath deposition (CBD) technique through controlled ionic reactions in an alkaline medium. Cadmium chloride monohydrate, manganese acetate tetrahydrate, and nickel chloride hexahydrate dissociate in aqueous solution to release $\mathrm{Cd^{2+}}$, $\mathrm{Mn^{2+}}$ and $\mathrm{Ni^{2+}}$ ions, respectively:
\begin{equation}
	\mathrm{CdCl_2\cdot H_2O}\rightarrow\mathrm{Cd^{2+}}+2\mathrm{Cl^-}+\mathrm{H_2O}
\end{equation}
\begin{equation}
	\mathrm{Mn(CH_3COO)_2\cdot 4H_2O}\rightarrow\mathrm{Mn^{2+}}+2\mathrm{CH_3COO^-}+4\mathrm{H_2O}
\end{equation}
\begin{equation}
	\mathrm{NiCl_2\cdot 6H_2O}\rightarrow\mathrm{Ni^{2+}}+2\mathrm{Cl^-}+6\mathrm{H_2O}
\end{equation}
Ammonia and triethanolamine (TEA) act as complexing agents, forming stable amine complexes with metal ions and thereby regulating their release:
\begin{equation}
	\mathrm{Cd^{2+}}+\mathrm{4NH_3}\rightleftharpoons\mathrm{[Cd(NH_3)_4]^{2+}}
\end{equation}
\begin{equation}
	\mathrm{Mn^{2+}}+\mathrm{6NH_3}\rightleftharpoons\mathrm{[Mn(NH_3)_6]^{2+}}
\end{equation}
\begin{equation}
	\mathrm{Ni^{2+}}+\mathrm{6NH_3}\rightleftharpoons\mathrm{[Ni(NH_3)_6]^{2+}}
\end{equation}
Thiourea serves as the sulfur source and undergoes alkaline hydrolysis to generate sulfide ions:
\begin{equation}
	\mathrm{(CSNH_2)_2}+2\mathrm{OH^-}\rightarrow\mathrm{CN_2H_2}+\mathrm{H_2S}+2\mathrm{H_2O}
\end{equation}
\begin{equation}
	\mathrm{H_2S}\rightleftharpoons\mathrm{S^{2-}}+2\mathrm{H^+}
\end{equation}
Subsequently, the controlled release of metal ions reacts with sulfide ions at the substrate interface, leading to the formation of Ni-doped $\mathrm{Cd_{1-x}}\mathrm{Mn_x}\mathrm{S}$ thin films:
\begin{equation}
	\mathrm{[Cd(NH_3)_4]^{2+}}+\mathrm{{[Mn(NH_3)_6]}^{2+}}+\mathrm{{[Ni(NH_3)_6]}^{2+}}+\mathrm{S^{2-}}\rightarrow\mathrm{Ni:}\mathrm{Cd_{1-x}}\mathrm{Mn_x}\mathrm{S}+\text{by products}
\end{equation}
This controlled reaction mechanism ensures uniform nucleation and growth of the thin film.
\begin{table}[ht]
	\centering
	\caption{Chemical bath composition and v/v \% of [Ni]/[Cd+Mn] for  Ni-doped $\mathrm{Cd_{1-x}}\mathrm{Mn_x}\mathrm{S}$ thin films}
	\label{tab:1}
	\resizebox{\textwidth}{!}{
		\begin{tabular}{l l c  c c c}
			\toprule
			Sample Code & CdCl$_2\cdot$H$_2$O (mL) & Mn(CH$_3$COO)$_2\cdot$H$_2$O (mL) & NH$_2$CSNH$_2$ (mL) & NiCl$_2\cdot$6H$_2$O (mL) & [Ni]/[Cd+Mn] (v/v \%) \\
			\midrule
			D1  & 6.0 & 4.0 & 20.0 & 0.1 & 1.0 \\
			D2  & 6.0 & 4.0 & 20.0 & 0.2 & 2.0 \\
			D3  & 6.0 & 4.0 & 20.0 & 0.3 & 3.0 \\
			D4  & 6.0 & 4.0 & 20.0 & 0.4 & 4.0 \\
			\bottomrule
		\end{tabular}
	}
\end{table}
\subsection{Characterization details}
Various sophisticated instruments were employed to investigate the structural, morphological, optical, and electrical properties of the Ni-doped $\mathrm{Cd_{1-x}Mn_xS}$ thin films. X-ray diffraction (XRD) analysis was performed to examine the structural characteristics using an X-ray diffractometer (Model: Empyrean, PANalytical) equipped with monochromatic $\mathrm{Cu-K}\upalpha$ radiation ($1.5406~\mathrm{\mathring{A}}$) operated at a maximum potential of 40 kV and maximum current of  100 mA. The measurements were carried out over a 2$\uptheta$ range of $10^\circ$ to $80^\circ$, with a step size of $0.026^\circ$. The crystallite size, microstrain, percentage of crystallinity, and other crystallographic parameters were determined from this XRD analysis. The surface morphology and film thickness of the deposited thin films were characterized using a Field-Emission Scanning Electron Microscope (FESEM, Model: ZEISS SIGMA) operated at an accelerating voltage of 20 kV. High-Resolution Transmission Electron Microscopy (HRTEM, Model: JEOL JEM-2100 Plus) operating at an accelerating voltage of 200 kV was employed to investigate the structural properties of the undoped and Ni-doped $\mathrm{Cd_{1-x}Mn_xS}$ thin films. Selected Area Electron Diffraction (SAED) patterns were recorded to identify the crystalline phase and determine interplanar spacing. An Energy-Dispersive X-ray (EDX) spectrometer attached to the HRTEM system was used to confirm the elemental composition of the samples. The optical properties of the thin films were investigated using a UV-Vis spectrophotometer (Model: Shimadzu UV-2600) in the wavelength range of 300–1100 nm. The absorbance and transmittance spectra were recorded and analysed using UVProbe software. The absorption coefficient ($\upalpha$), optical band gap ($\mathrm{E_g}$), and other optical parameters, including refractive index and extinction coefficient, were derived from the measured spectra. Polarity and the Hall parameters  of the thin films were determined using a Hall measurement system (Model: OMEGA, EMU-10) in the Van der Pauw configuration. The current–voltage (I–V) characteristics of the thin films were measured using a source meter (Model: Keithley 2450) integrated with a standard four-probe configuration. The electrical measurements were controlled and recorded using KickStart software. A calibrated solar simulator equipped with a xenon lamp was employed to assess the electrical performance under standard illumination conditions. The pH of the precursor solutions was monitored using a digital pH meter with a resolution of 0.01. \\The prepared samples were designated as D1, D2, D3, and D4, corresponding to 1\%, 2\%, 3\%, and 4\% Ni-doped $\mathrm{Cd_{1-x}Mn_xS}$ thin films, respectively. It was observed that Ni doping beyond 4\% led to noticeable changes in the crystal structure and the emergence of secondary phases. Therefore, higher doping concentrations were avoided to maintain phase purity and structural stability. The Ni-doped $\mathrm{Cd_{1-x}}\mathrm{Mn_x}\mathrm{S}$ thin films, exhibit a characteristic yellow colour, with transparency gradually decreasing as the Ni doping concentration increases.
\section{Results and discussion}
\subsection{Structural analysis}
\begin{figure}[ht]
	\centering
	\includegraphics[width=0.8\textwidth,height=0.4\textheight]{"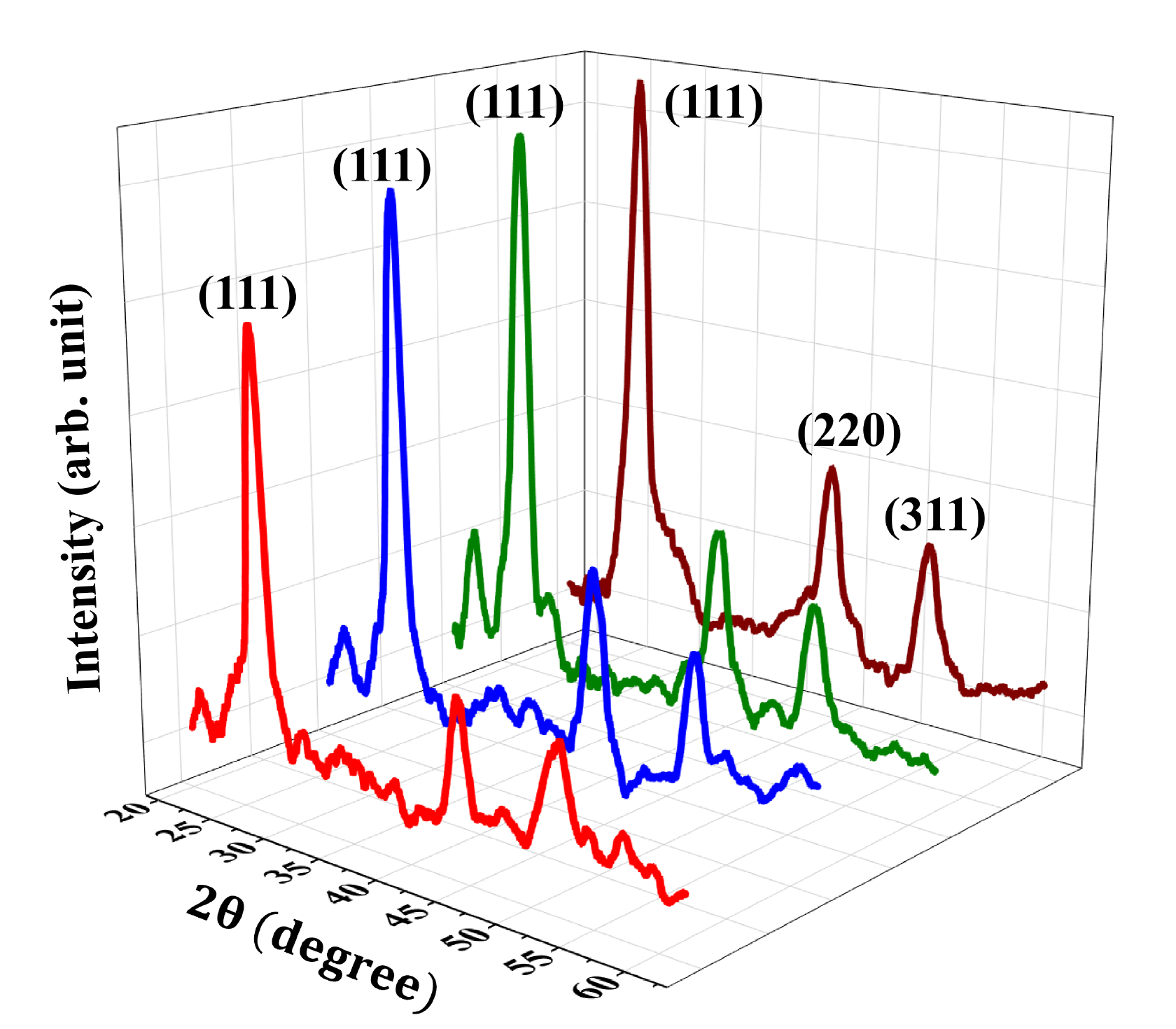"}
	\caption{XRD pattern of Ni doped $\mathrm{Cd_{1-x}Mn_xS}$ thin films}
	\label{fig:XRD}
\end{figure}

\begin{figure}[ht]
	\centering
	\includegraphics[width=0.8\textwidth,height=0.4\textheight]{"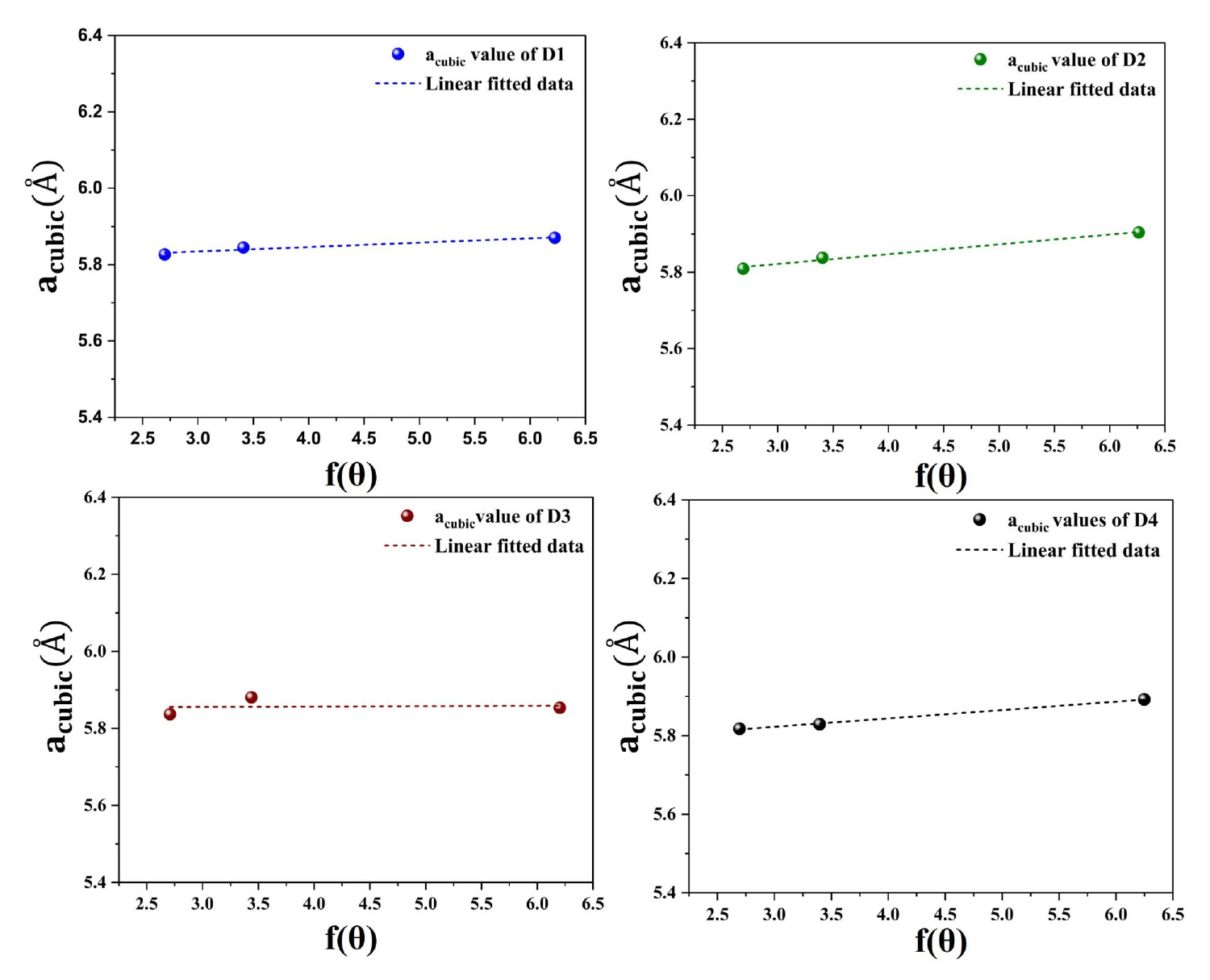"}
	\caption{N-R plot for lattice constant calculation of Ni doped $\mathrm{Cd_{1-x}Mn_xS}$ thin films}
	\label{fig:Lattice_cosntant}
\end{figure}

\begin{figure}[ht]
	\centering
	\includegraphics[width=0.8\textwidth,height=0.4\textheight]{"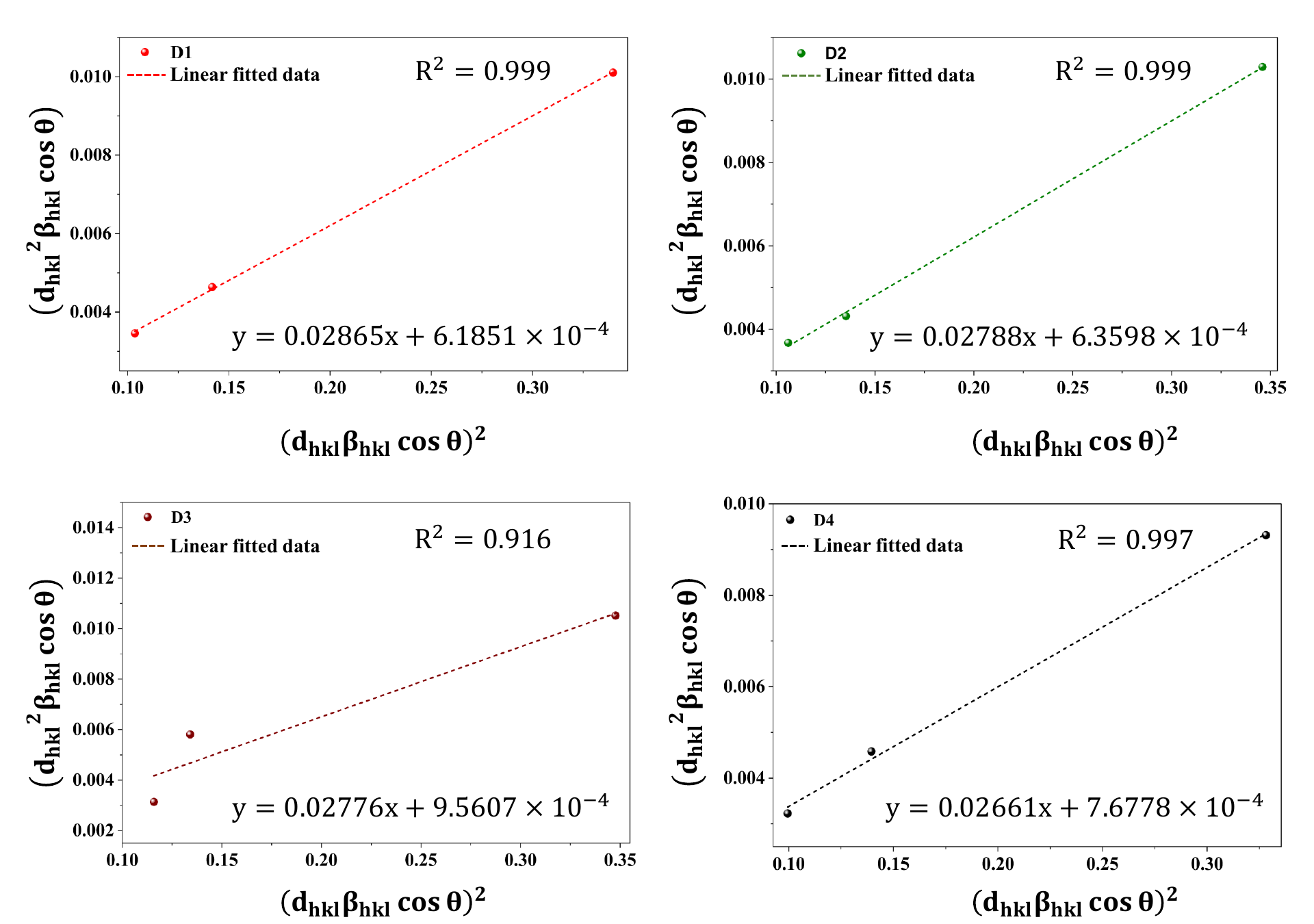"}
	\caption{Size-strain plot of Ni doped $\mathrm{Cd_{1-x}Mn_xS}$ thin films}
	\label{fig:SS_plot}
\end{figure}

Figure \ref{fig:XRD} presents the X-ray diffraction (XRD) patterns of the deposited thin films. CdS and MnS are known to exist in several polymorphic forms. CdS generally crystallizes in both cubic and hexagonal phases [8]. In contrast, MnS occurs in three different crystalline forms: $\upalpha$-MnS (rock salt structure), $\upbeta$-MnS (zinc blende structure), and $\upgamma$-MnS (wurtzite structure) \cite{hannachi2016preparation}. The obtained diffraction peaks of the $\mathrm{Cd_{1-x}}\mathrm{Mn_x}\mathrm{S}$ thin films were found to lie close to the standard peak positions of cubic CdS (JCPDS card no. 75-1546) and cubic MnS (JCPDS card no. 76-2052). Comparison of the experimental diffraction data with the standard JCPDS card for $\mathrm{Cd_{1-x}}\mathrm{Mn_x}\mathrm{S}$ (card no. 01-083-5259) indicates that the deposited $\mathrm{Cd_{1-x}}\mathrm{Mn_x}\mathrm{S}$ thin films crystallize in a cubic zinc blende structure. The observed peak broadening suggests the formation of nanocrystalline thin films \cite{boruah2025low}. 
No additional diffraction peaks were observed with increasing Ni doping concentration up to a certain limit, indicating that the incorporation of Ni ions does not lead to the formation of secondary phases. A slight shift of the peaks found in the thin films. The position of the prominent (111) diffraction peak initially shifted towards lower angles from 26.273° for the 1\% Ni-doped sample to 26.121° for the 2\% Ni-doped sample, which may be attributed to a slight lattice expansion associated with the substitutional incorporation of Ni ions into the $\mathrm{Cd_{1-x}Mn_xS}$ lattice. With further increase in Ni concentration, the peak shifted towards higher angles to 26.351° for the 3\% Ni-doped sample and subsequently shifted back to a lower angle of 26.175° for the 4\% Ni-doped sample. The absence of a monotonic peak shift indicates that the incorporation of Ni ions gives rise to local lattice distortions and strain fluctuations within the host lattice. Such non-uniform variations in peak position may result from the combined effects of substitutional incorporation, defect formation, and strain relaxation mechanisms \cite{Gbashi, Goktas}. Instead, an increase in the intensity of the diffraction peaks is observed, which is clearly evident for the prominent (111) peak in the three-dimensional XRD pattern, suggesting enhanced crystallinity and crystallite size of the deposited films \cite{salem2020fabrication,pathok2026comprehensive}.\\The crystallite size (D) of the deposited thin films was determined using the Scherrer equation, as given in Eq. \ref{eq:crystallite_size} \cite{kumar2012cds, Najim}:
\begin{equation}
	\mathrm{D} = \frac{\mathrm{K}\uplambda_\mathrm{x}}{\upbeta_{\mathrm D} \cos\uptheta}
	\label{eq:crystallite_size}
\end{equation}
where D is the crystallite size, K is the shape factor (0.94 for spherical particles), $\uplambda_x$ is the X-ray wavelength, $\upbeta_{\mathrm{D}}$ is the FWHM in radians, and $\uptheta$ is the Bragg angle. The average crystallite size of the $\mathrm{Cd_{1-x}}\mathrm{Mn_x}\mathrm{S}$ thin films increases from 4.50 nm to 4.66 nm with increasing Ni doping concentration. The values were obtained by averaging the crystallite sizes from the (111), (220), and (311) planes, and are listed in Table \ref{tab:Crystallite_size_Lattice__constant}. This trend suggests that Ni doping influences the microstructural evolution of the films. A similar observation was reported by K. Chaitanya Kumar et al. for Ni-doped ZnS thin films \cite{kumar2019effect}. \\Interplanar spacing ($\mathrm{d_{hkl}}$) of the films was calculated by using Bragg’s relation, Eq. \ref{eq:interplanar_spacing} \cite{ashith2023influence}:
\begin{equation}
	\mathrm{d_{hkl}}=\frac{\uplambda_\mathrm{x}}{2\sin\uptheta}
	\label{eq:interplanar_spacing}
\end{equation}
Where $\uplambda_x$ is the wavelength of the X-ray and $\uptheta$ is the Bragg angle. The interplanar spacing was found to be in the range of 3.38–3.41 $\mathring{\mathrm{A}}$ for the (111) plane, 2.06–2.08 $\mathring{\mathrm{A}}$ for the (220) plane, and 1.75–1.76 $\mathring{\mathrm{A}}$ for the (311) plane, as presented in Table \ref{tab:Structural_parameters}.
\begin{table}[ht]
	\centering
	\renewcommand{\arraystretch}{1.3}
	\setlength{\tabcolsep}{43pt}
	\caption{X-ray diffraction parameters of Ni-doped $\mathrm{Cd_{1-x}Mn_xS}$ thin films}
	\label{tab:Structural_parameters}
	\begin{tabularx}{\linewidth}{c c  c c}
		\hline
		Sample & \makecell{2$\uptheta$ (deg.)} & (hkl) &  \makecell{$\mathrm{d_{hkl}}$ ($\mathring{\mathrm{A}}$)}\\
		\hline
		& 26.273 & (111)  & 3.39 \\
		D1 & 43.771 & (220)  & 2.07\\
		& 52.013 & (311)  & 1.76\\
		& 26.121 & (111)  & 3.41 \\
		D2 & 43.827 & (220)  & 2.06\\
		& 52.183 & (311)  & 1.75\\
		& 26.351 & (111)  & 3.38 \\
		D3 & 43.493 & (220)  & 2.08\\
		& 51.917 & (311)  & 1.76\\
		& 26.175 & (111)  & 3.40 \\
		D4 & 43.897 & (220)  & 2.06\\
		& 52.101 & (311)  & 1.75\\
		\hline
	\end{tabularx}
\end{table}
The lattice constant ($\mathrm{a_{cubic}}$) of films was calculated by using Eq.\ref{eq:a_cubic} \cite{kumar2012cds}.
\begin{equation}
	\mathrm{a_{cubic}}=\mathrm{d_{hkl}}\sqrt{\mathrm{h^2+k^2+l^2}}
	\label{eq:a_cubic}
\end{equation}
Where h, k, and l are the Miller indices and $\mathrm{d_{hkl}}$ is the interplanar spacing. The lattice constant of the thin films was found to range from 5.8089 $\mathring{\mathrm{A}}$ to 5.9041 $\mathring{\mathrm{A}}$, as presented in Table \ref{tab:Crystallite_size_Lattice__constant}. The changes in lattice constant with Ni doping can be attributed to the substitution of Ni ions into the $\mathrm{Cd_{1-x}Mn_xS}$ lattice, which induces lattice distortion and modifies the interplanar spacing, as evidenced by the observed shifts in the XRD peaks.\\In XRD analysis, the accuracy of the measured diffraction angle ($\uptheta$) and the corresponding interplanar spacing ($\mathrm{d_{hkl}}$) is influenced by several systematic factors, including divergence of the incident X-ray beam and absorption or reflection effects within the specimen. The interplanar spacing is expressed as $\mathrm{d_{hkl}}=\frac{\uplambda_x}{2}\csc\uptheta$, while the associated error is given by $\updelta\mathrm{d_{hkl}}=\frac{-\uplambda_x}{2}(\csc\uptheta\cot\uptheta)\updelta\uptheta$. It is evident that as $\uptheta$ approaches $90^\circ$, the ratio $\frac{\mathrm{d_{hkl}}}{\updelta\mathrm{d_{hkl}}}$ tends to zero, indicating a significant reduction in measurement error at higher diffraction angles. To minimize these systematic errors and obtain a reliable lattice parameter, the Nelson–Riley method is employed. In this method, the calculated lattice parameters are plotted against the error function $\mathrm{f}(\uptheta)=\frac{1}{2}\left[\left(\frac{\cos^2\uptheta}{\sin\uptheta}\right)+\left(\frac{\cos^2\uptheta}{\uptheta}\right)\right]$, and the corrected lattice constant is obtained by extrapolating the linear fit to $\mathrm{f}(\uptheta)=0$, as shown in Figure \ref{fig:Lattice_cosntant}.
The corrected lattice constants, as presented in Table \ref{tab:Crystallite_size_Lattice__constant}, indicate a systematic variation with increasing Ni doping concentration. \\The crystallite size and microstrain of the thin films were calculated using the Size-Strain Plot (SSP) method. In this method, the XRD peak profile is considered a combination of Lorentzian and Gaussian functions. The size-broadening component of the XRD profile is represented by a Lorentzian function, whereas the strain-broadening component is represented by a Gaussian function. The SSP method generally gives better results for isotropic broadening because it gives more weight to low-angle reflections, where accuracy and precision are higher than at higher angles. This is because XRD data at higher angles are of lower quality, and the peaks are often highly overlapped at higher diffraction angles. \\Therefore, the overall peak broadening in the SSP method can be expressed as follows:
\begin{equation}
	\upbeta_{\mathrm{hkl}}=\upbeta_\mathrm{L}+\upbeta_\mathrm{G}
	\label{eq:FWHM_SSP}
\end{equation}
where, $\upbeta_\mathrm{L}$ and $\upbeta_\mathrm{G}$ represents the peak broadening due to Lorentz and Gaussian function, respectively. \\Using the Size–Strain Plot (SSP) method, the crystallite size (D) and microstrain ($\upvarepsilon$) of the deposited thin films were calculated from Eq. \ref{eq:SSP} \cite{Zak}.
\begin{equation}
	(\mathrm{d_{hkl}\upbeta_{hkl}\cos\uptheta})^2=\frac{\mathrm{k}\uplambda_\mathrm{x}}{\mathrm{D}}\mathrm{d_{hkl}^2\upbeta_{hkl}\cos\uptheta}+\frac{\upvarepsilon^2}{4}
	\label{eq:SSP}
\end{equation}
where $\mathrm{d_{hkl}}$ denotes the interplanar spacing of the (hkl) plane and $\uptheta$ is the Bragg angle. The parameter k is the shape factor, taken as 0.94 and $\uplambda_\mathrm{x}$ is the wavelength of the incident X-ray radiation.\\ The SSP plot was constructed by plotting $\mathrm{d_{hkl}^2\upbeta_{hkl}\cos\uptheta}$ on the X-axis and $(\mathrm{d_{hkl}\upbeta_{hkl}\cos\uptheta})^2$ on the Y-axis for each diffraction peak, as shown in Figure \ref{fig:SS_plot}. The crystallite size was determined from the slope of the linear fit, whereas the lattice microstrain was obtained from the intercept. The calculated values of crystallite size and microstrain for the prepared thin films are summarized in Table \ref{tab:Crystallite_size_Lattice__constant}. The crystallite sizes obtained from both the Scherrer equation and SSP method follow the same trend, showing an increase with increasing Ni concentration in the $\mathrm{Cd_{1-x}Mn_xS}$ thin films. However, a slight difference in the estimated values is observed, which may be attributed to the consideration of lattice strain effects in the SSP method, whereas the Scherrer equation accounts only for size-induced broadening \cite{Boro}. A similar phenomenon has been reported by D. Nath et al. during the estimation of crystallite size in CdSe thin films \cite{Nath}.

The dislocation density ($\updelta$) of the deposited films were calculated by Eq. \ref{eq:dislocation} \cite{fouad2018multifunctional},
\begin{table}[ht]
	\centering
	\renewcommand{\arraystretch}{1.8}
	\setlength{\tabcolsep}{3.6 pt}
	\caption{Details of estimated average crystallite size and Lattice constant for Ni-doped $\mathrm{Cd_{1-x}Mn_xS}$ thin films}
	\label{tab:Crystallite_size_Lattice__constant}
	\begin{tabularx}{\linewidth}{c c c c c c}
		\hline
		\makecell{\text{Sample}\\\text{Code}} & 
		\multicolumn{2}{c}{Average crystallite size (D)} & \multicolumn{2}{c}{\text{Lattice constant ($\mathrm{a_{cubic}}$)}} & \makecell{Microstrain ($\upvarepsilon$)\\($\times~10^{-3}$)}\\
		\cline{2-5}
		& \makecell{Scherrer's formula\\(nm)} & \makecell{SSP method\\(nm)} & \makecell{Lattice constant\\($\mathring{\mathrm{A}}$)} & \makecell{Corrected lattice constant\\($\mathring{\mathrm{A}}$)} & \\
		\hline
		&      &      & 5.8705 &     &  \\
		D1 & 4.50 & 5.05 & 5.8450 & 5.8001 & 49.74\\
		&      &      & 5.8266 &      & \\
		&      &      & 5.9041 &      & \\
		D2 & 4.53 & 5.19 & 5.8379 & 5.7448 & 50.43\\
		&      &      & 5.8089 &       &\\
		&      &      & 5.8534 &       &\\
		D3 & 4.54 & 5.22 & 5.8805 & 5.8536 & 62.84\\
		&      &      & 5.8366 &       & \\
		&      &      & 5.8921 &       & \\
		D4 & 4.66 & 5.44 & 5.8290 & 5.7585 & 55.42\\
		&      &      & 5.8174 &      & \\
		\hline
	\end{tabularx}
\end{table} 
\begin{equation}
	\updelta=\frac{1}{\mathrm{D^2}}
	\label{eq:dislocation}
\end{equation}
where D is the crystallite size. The dislocation density of the $\mathrm{Cd_{1-x}Mn_xS}$ thin films was found to decrease from 0.0290 $\mathrm{nm^{-2}}$ to 0.0137 $\mathrm{nm^{-2}}$ with increasing Ni doping concentration (Table \ref{tab:dislocation_crystallinity})\\The crystallinity of the Ni-doped thin films was calculated by using Eq. \ref{eq:crystallinity} \cite{guler2019annealing}:
\begin{equation}
	\mathrm{Crystallinity}\%=\frac{\text{Total area of the crystalline peaks}}{\text{Total area under all peaks}}\times 100
	\label{eq:crystallinity}
\end{equation}
Where the total area under all peaks represents the sum of both crystalline and amorphous contributions, including broad humps or halos in the XRD pattern. The crystallinity percentages were found to be 77\%, 79\%, 81\%, and 83\% for the D1, D2, D3, and D4 thin film samples, respectively,  as summarized in Table \ref{tab:dislocation_crystallinity}. It was observed that the crystallinity of the films improved progressively with increasing Ni doping concentration. This improvement in crystallinity is crucial for achieving a high-quality window layer in thin-film solar cells (TFSCs), as reported by Tsuji et al. \cite{tsuji2000characterization}.\\The residual stress ($\upsigma_s$) in the deposited D1–D4 thin films was determined using Eq. \ref{eq:residual_stress} \cite{guler2019annealing,kumari2022interrelation}:
\begin{equation}
	\upsigma_s=-45.82\times 10^9\left(\frac{\mathrm{a-a_0}}{\mathrm a_0}\right)
	\label{eq:residual_stress}
\end{equation}
where a and $\mathrm{a_0}$ represent the experimental and standard lattice parameters, respectively. The standard lattice parameter $\mathrm{a_0}$ = 5.8290 $\mathring{\mathrm{A}}$ was obtained from the JCPDS card corresponding to $\mathrm{Cd_{1-x}Mn_xS}$ (x=0.4) (Card No. 01-083-5259) and biaxial modulus $45.82\times10^9$ Pa was derived from the elastic stiffness constants C11 and C12 as reported by A. Rani et al. \cite{rani2015stability}. As shown in Table 4, the residual stress values of Ni-doped $\mathrm{Cd_{1-x}Mn_xS}$ thin films range from –0.1342 GPa to –0.2176 GPa, with the negative sign indicating compressive stress within the crystal structure. This compressive nature arises from lattice contraction induced by Ni$^{2+}$ incorporation, as the ionic radius of $\mathrm{Ni^{2+}}$ (0.55 $\mathring{\mathrm{A}}$) is smaller than that of $\mathrm{Cd^{2+}}$ (0.97 $\mathring{\mathrm{A}}$) and $\mathrm{Mn^{2+}}$ (0.83 $\mathring{\mathrm{A}}$) in the host lattice \cite{Ashokkumar, Khallaf, D'Silva}. However, the residual stress does not vary monotonically with increasing Ni doping concentration, despite the monotonic improvement in crystallinity and the corresponding reduction in strain \cite{Demircan}. The residual stress progressively increases from – 0.1434 GPa for sample D1 to – 0.1663 GPa for D2 and reaches its maximum compressive value of – 0.2176 GPa for D3, but then decreases to – 0.1342 GPa at the highest doping concentration D4. This non-monotonic behavior reflects the competition between two opposing mechanisms: dopant-induced lattice contraction increases compressive stress, while microstructural improvements including enhanced crystallinity and reduced strain relax internal stresses by minimizing defect density and grain boundary contributions \cite{Kaphle,Liu}. At lower Ni concentrations from D1 to D3, lattice contraction dominates, leading to increasing compressive stress. At the highest concentration D4, the superior crystallinity and minimized strain counteract the lattice contraction effect, resulting in reduced compressive stress. Similar non-monotonic stress trends have been reported in the literature, such as by Kaphle et al. for Co-doped ZnO thin films, confirming that multiple competing stress-contributing factors can yield non-monotonic behavior even when microstructural parameters evolve monotonically\cite{Kaphle}.
\begin{table}[ht]
	\centering
	\renewcommand{\arraystretch}{1.3}
	\setlength{\tabcolsep}{24pt}
	\caption{Dislocation density, crystallinity, and residual stress of Ni-doped $\mathrm{Cd_{1-x}Mn_xS}$ thin films}
	\label{tab:dislocation_crystallinity}
	\begin{tabularx}{\linewidth}{c c c c}
		\hline
		Sample & $\updelta$ (nm$^{-2}$) & Crystallinity (\%) & Residual stress (GPa) \\
		\hline
		D1 & 0.0290 & 77 & -0.1434 \\
		D2 & 0.0274 & 79 & -0.1663 \\
		D3 & 0.0201 & 81 & -0.2176 \\
		D4 & 0.0137 & 83 & -0.1342 \\
		\hline
	\end{tabularx}
\end{table}
\subsection{High Resolution Transmission Electron Microscopy analysis}
\begin{figure}[ht]
	\centering
	\includegraphics[width=0.8\textwidth,height=0.4\textheight]{"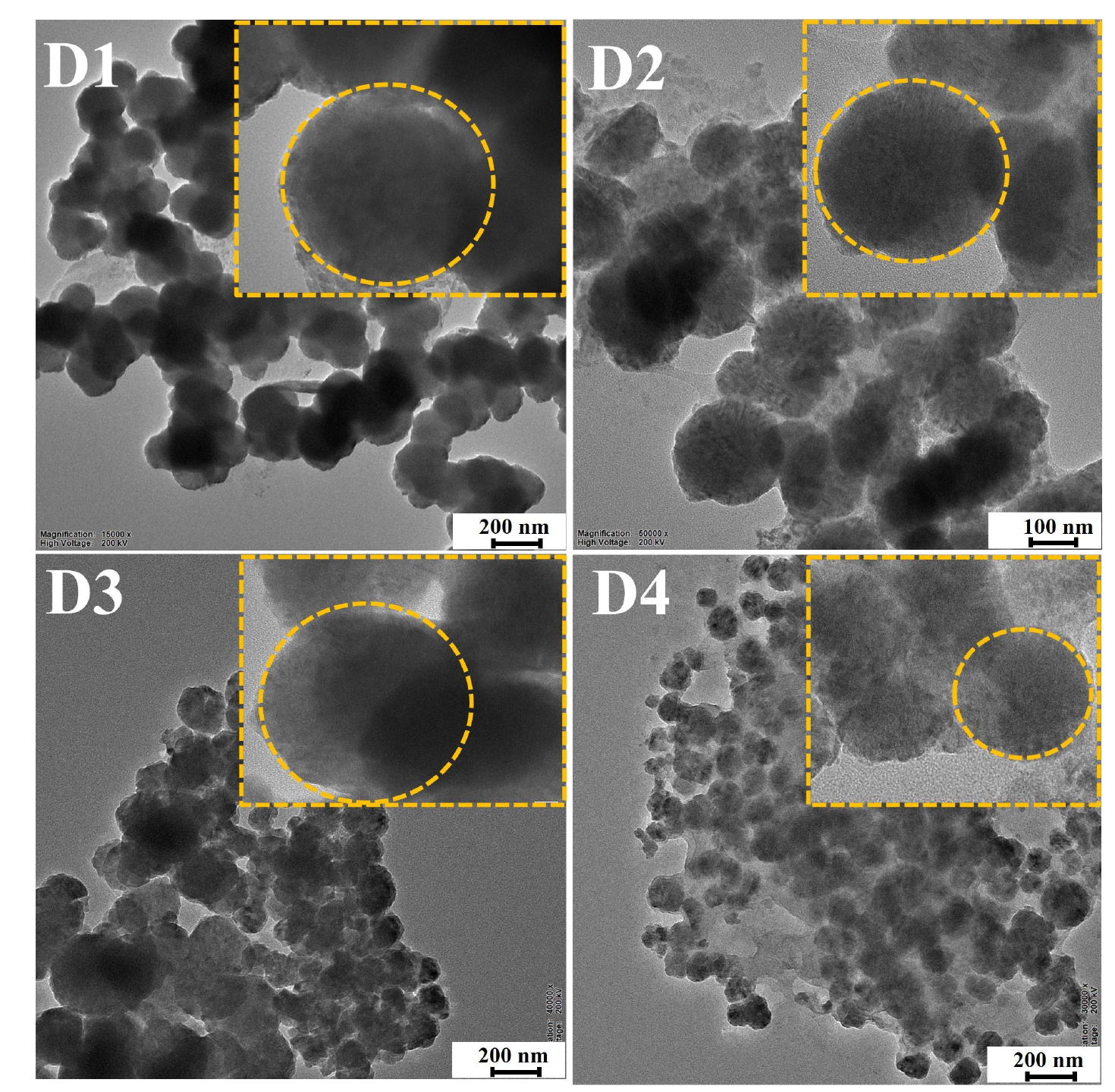"}
	\caption{HRTEM images of the spherical nanosized grains in Ni doped $\mathrm{Cd_{1-x}Mn_xS}$ thin films}
	\label{fig:Grain}
\end{figure}

\begin{figure}[ht]
	\centering
	\includegraphics[width=0.8\textwidth]{"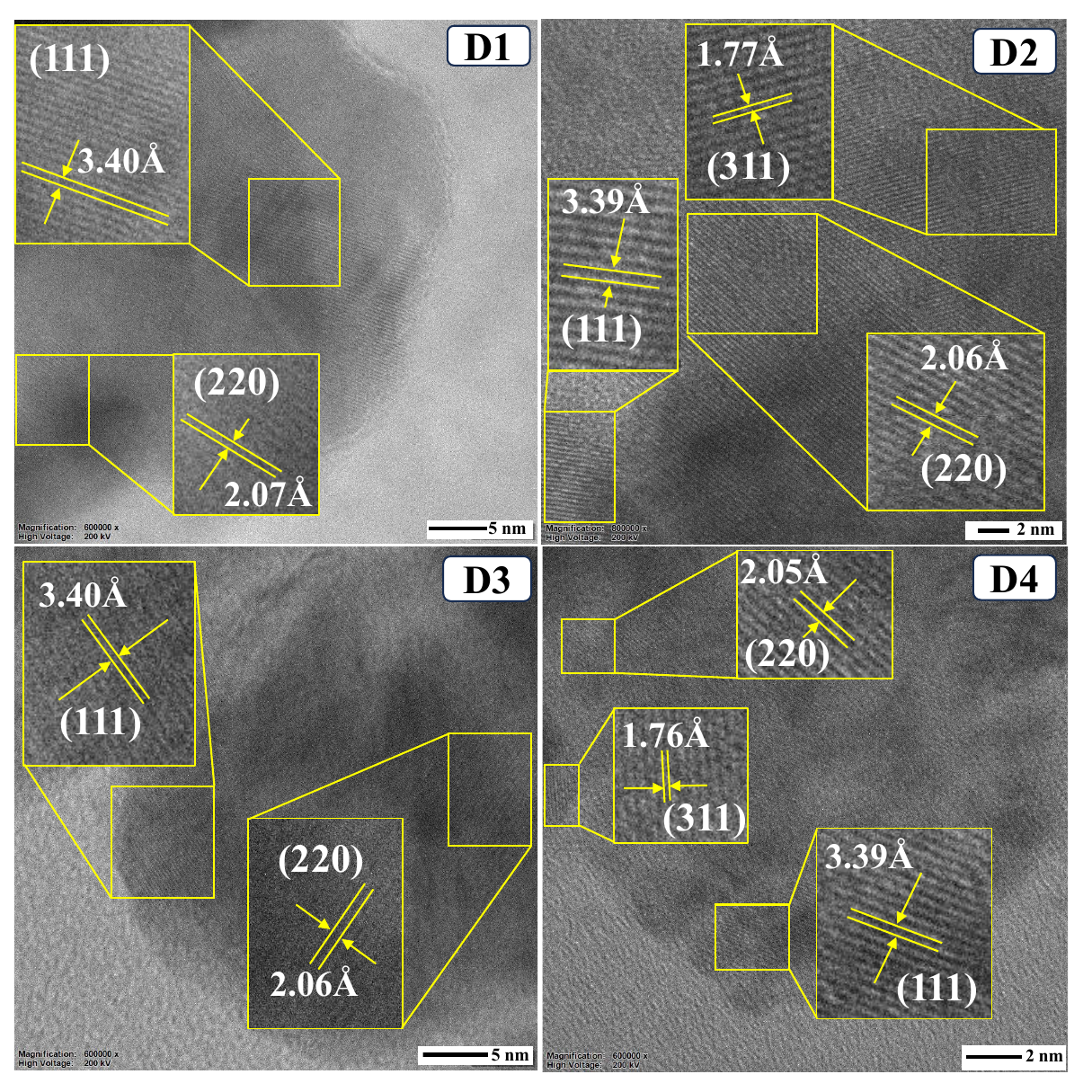"}
	\caption{HRTEM micrographs in Ni doped $\mathrm{Cd_{1-x}Mn_xS}$ thin films revealing crystalline lattice planes}
	\label{fig:interplanar_spacing}
\end{figure}

\begin{figure}[ht]
	\centering
	\includegraphics[width=0.8\textwidth,height=0.4\textheight]{"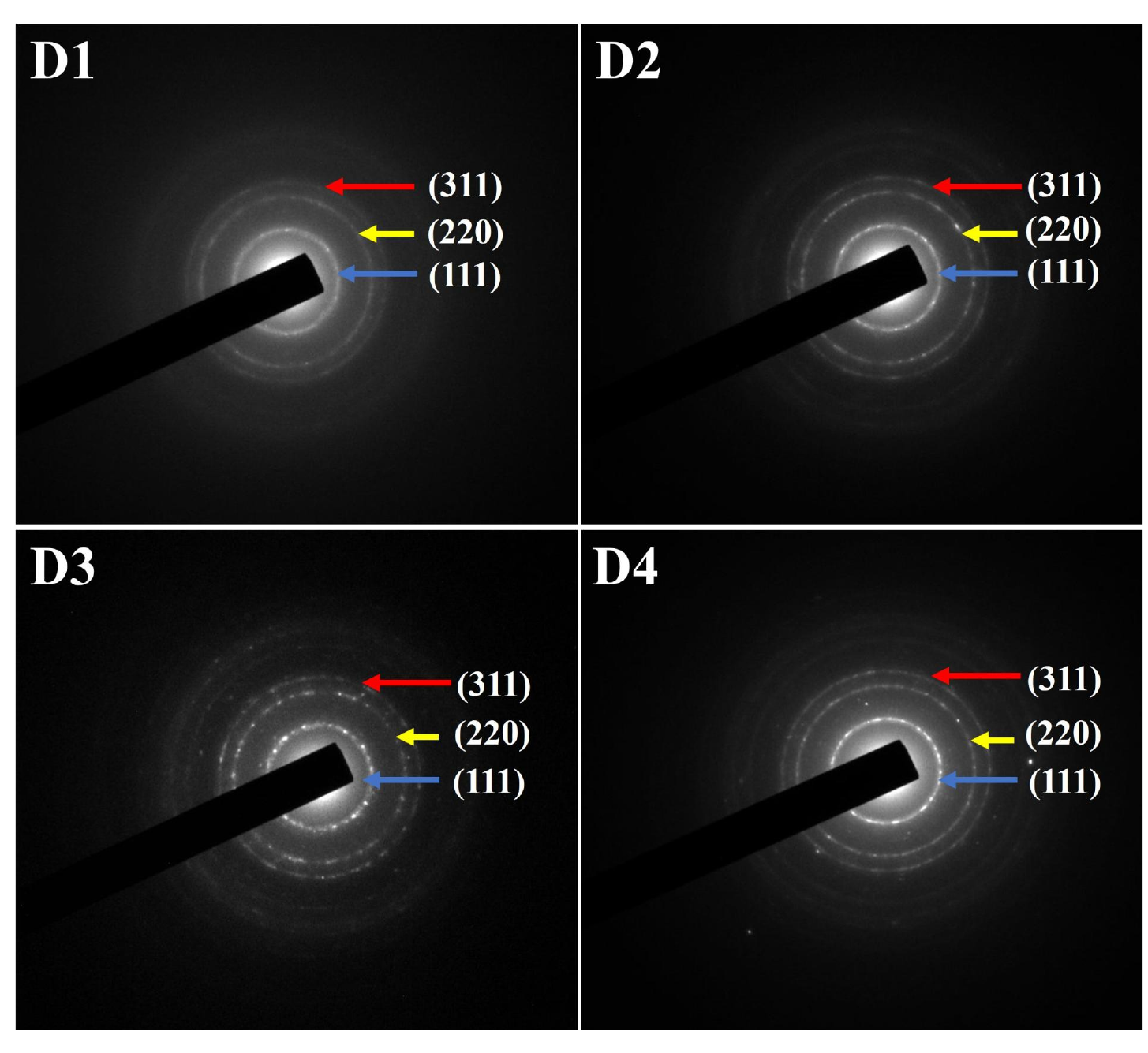"}
	\caption{SAED pattern of Ni doped $\mathrm{Cd_{1-x}Mn_xS}$ thin films}
	\label{fig:SAED}
\end{figure}

Figures \ref{fig:Grain} present the HRTEM micrographs of the Ni-doped thin films (D1-D4). The micrographs show that the grains are nearly spherical in shape. The average grain size, estimated from these micrographs by using ImageJ software and found to be increase from 90.01$\pm$0.69 nm to 147.90$\pm$1.31 nm as the doping concentration of Ni on the $\mathrm{Cd_{1-x}Mn_xS}$ thin film increases. The calculated average grain sizes are summarized in Table \ref{tab:EDX}. In addition, the high-resolution TEM images shown in Figure \ref{fig:interplanar_spacing} clearly reveal lattice fringes of the thin films. The interplanar spacing values estimated from HRTEM images by using ImageJ software were found to be in good agreement with those calculated from XRD analysis, confirming the reliability of the structural characterization and the polycrystalline nature of the films. The SAED pattern shown in Figure \ref{fig:SAED} exhibits concentric diffraction rings corresponding to different crystalline planes of the Ni-doped thin films. The appearance of continuous rings rather than discrete spots indicates that the films are polycrystalline in nature. The sharpness and well-defined character of the rings further suggest good crystallinity of the deposited films.
\begin{table}[ht]
	\centering
	\renewcommand{\arraystretch}{1.8}
	\setlength{\tabcolsep}{14pt}
	\caption{Atomic percentage (at.\%) composition of constituent elements, [Ni]/[Cd+Mn] ratios and average grain size of the Ni-doped $\mathrm{Cd_{1-x}Mn_xS}$ thin films}
	\label{tab:EDX}
	\begin{tabularx}{\linewidth}{c c c c c c c }
		\hline
		\makecell{Sample\\Code} & \makecell{Cd\\at.\%} & \makecell{Mn\\at.\%} & \makecell{S\\at.\%} & \makecell{Ni\\at.\%} & \makecell{$\mathrm{\frac{[Ni]}{[Cd+Mn]}}$(\%)} &\makecell{Average grain size\\(nm)}\\
		\hline		
		D1 & 29.9 & 19.9 & 49.7 & 0.5 & 1.0 & 90.01$\pm$0.69\\
		D2 & 29.4 & 19.6 & 50.0 & 1.0 & 2.0 & 114.32$\pm$0.72\\	
		D3 & 29.3 & 19.4 & 49.8 & 1.5 & 3.1 & 130.63$\pm$0.49\\
		D4 & 28.8 & 19.6 & 49.7 & 1.9 & 3.9 & 147.90$\pm$1.31\\
		\hline
	\end{tabularx}
\end{table} 

\begin{figure}[ht]
	\centering
	\includegraphics[width=0.8\textwidth,height=0.5\textheight]{"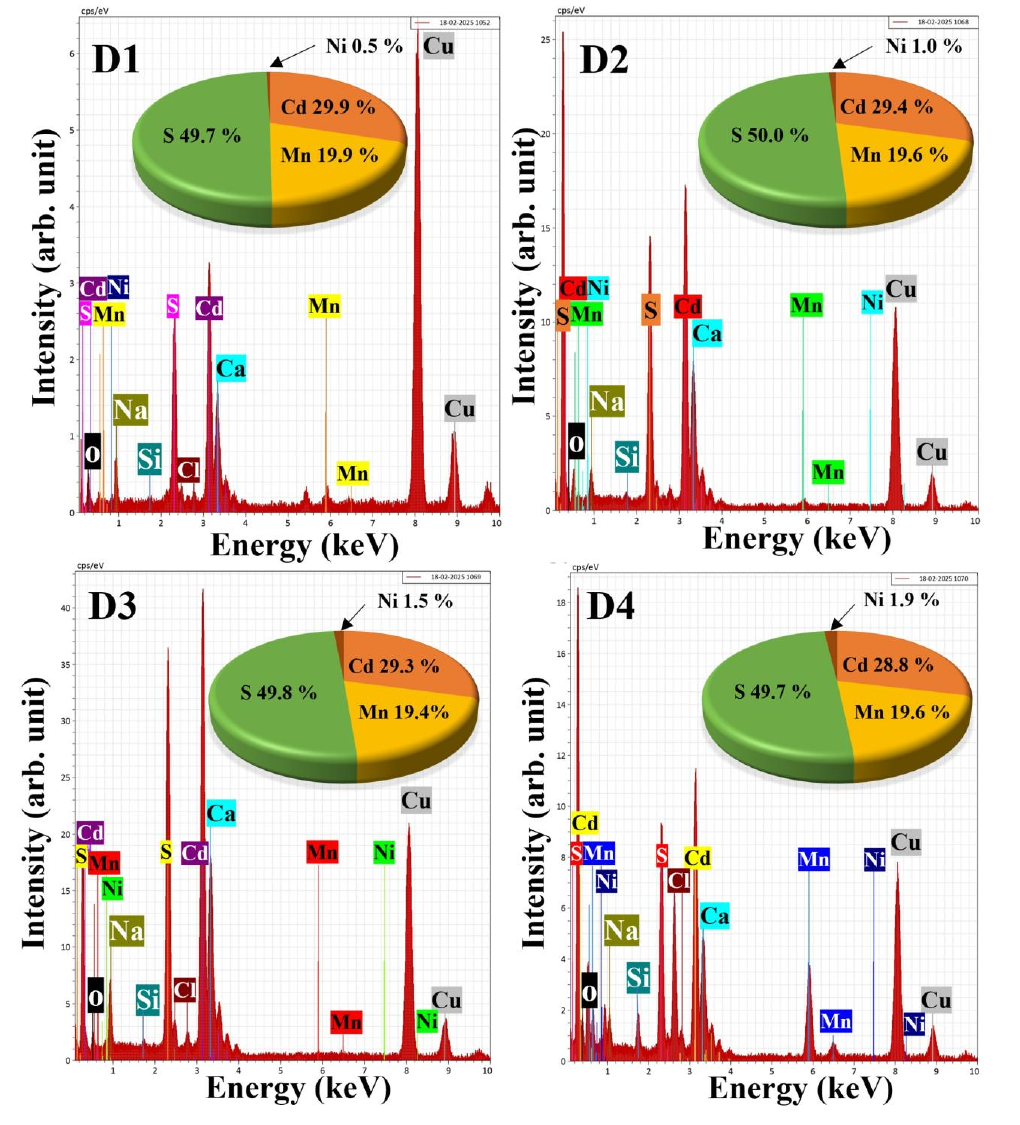"}
	\caption{EDX spectra of Ni doped $\mathrm{Cd_{1-x}Mn_xS}$ thin films}
	\label{fig:EDX}
\end{figure}

EDX analysis (Figure \ref{fig:EDX}) verified that  Cd, Mn, and S are present in all thin film samples, while the Ni content increases gradually with higher doping level. In addition to these elements, Ca and O peaks were observed, originating from the underlying glass substrate, while Cu signals originated from the HRTEM grid used during sample preparation. A weak Cl signal is evident, which can be linked to the precursor salts  ($\mathrm{CdCl_2.2H_2O}$ and $\mathrm{NiCl_2.6H_2O}$). These extraneous elements were excluded from the quantitative analysis, and Table \ref{tab:EDX} presents only the atomic percentages of Cd, Mn, S, and Ni. The observations confirm that the chosen deposition parameters ensured near-stoichiometric $\mathrm{Cd_{1-x}Mn_xS}$ thin films with controlled Ni incorporation.
\subsection{Morphological analysis}
\begin{figure}[ht]
	\centering
	\includegraphics[width=0.8\textwidth,height=0.4\textheight]{"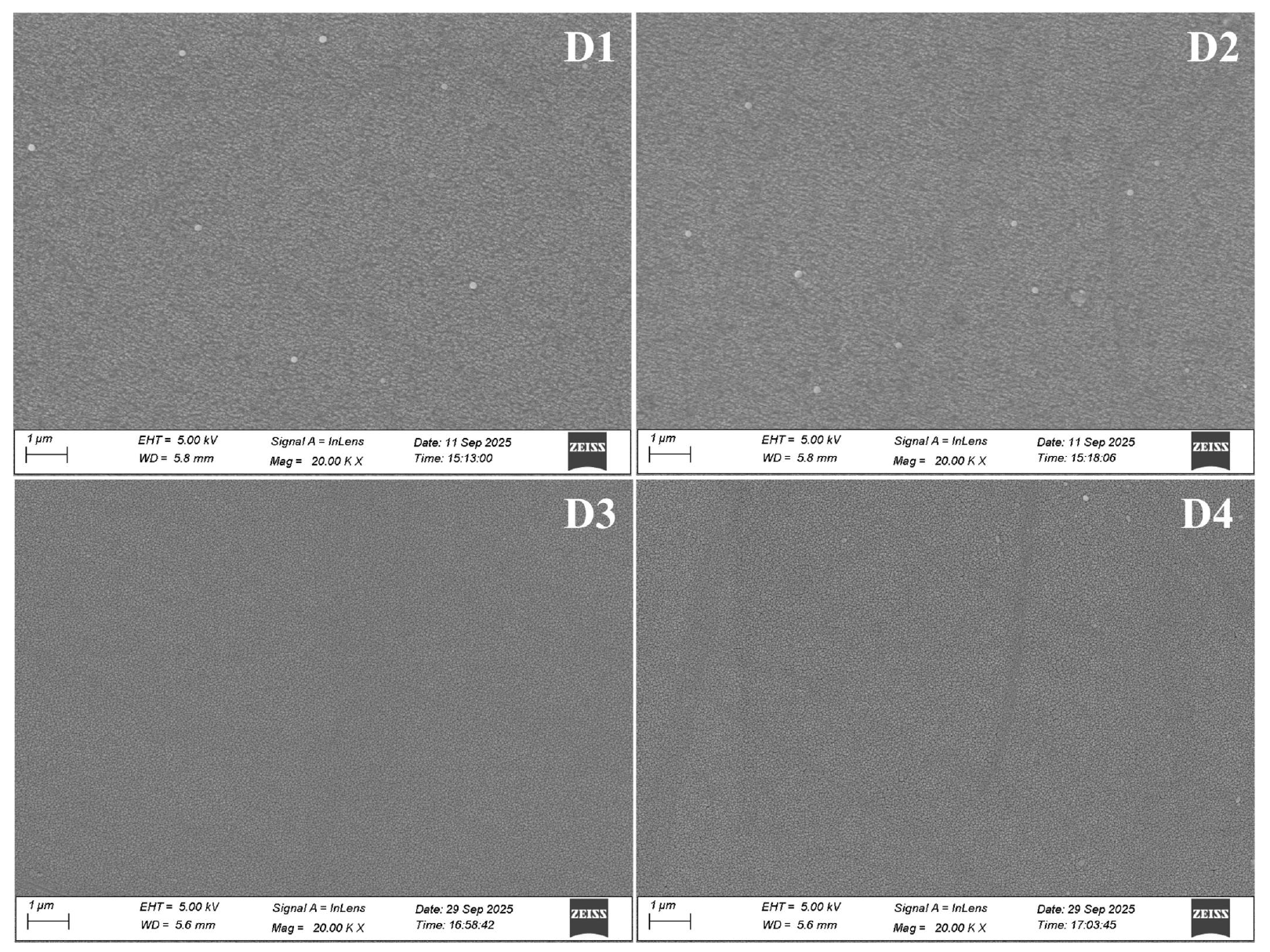"}
	\caption{FESEM images of Ni doped $\mathrm{Cd_{1-x}Mn_xS}$ thin films}
	\label{fig:FESEM}
\end{figure}

\begin{figure}[ht]
	\centering
	\includegraphics[width=0.8\textwidth,height=0.4\textheight]{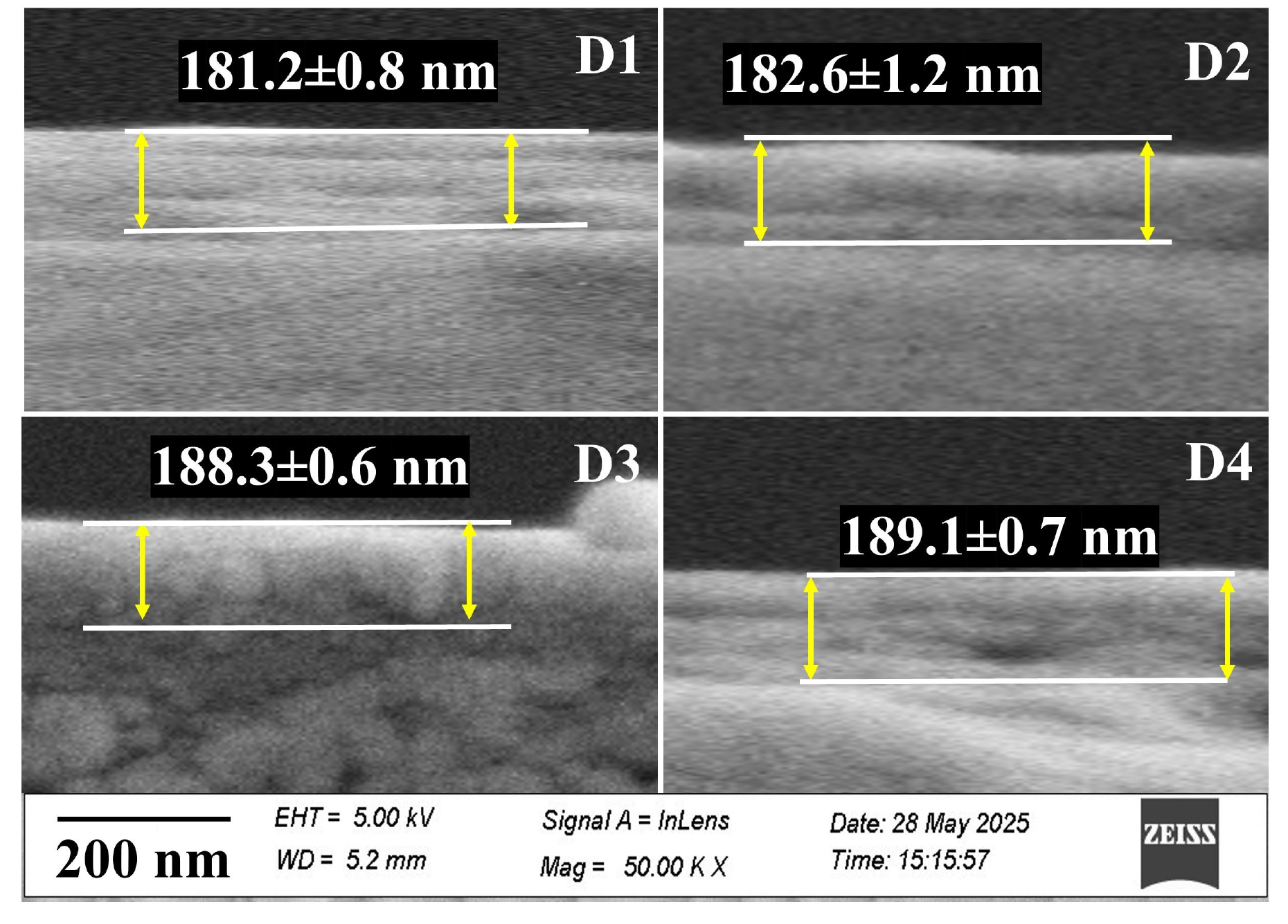}
	\caption{FESEM cross sectional images of Ni doped $\mathrm{Cd_{1-x}Mn_xS}$ thin films}
	\label{fig:thickness}
\end{figure}

Figure \ref{fig:FESEM} shows the FESEM micrographs of the Ni-doped $\mathrm{Cd_{1-x}Mn_xS}$ thin films. The images reveal a smooth and uniform surface morphology with almost no voids. The grains are densely packed and well interconnected, indicating compact film formation. No cracks are observed on the surface, indicating good film integrity. In addition, the films appear to be well adhered to the substrate, which suggests good mechanical stability and successful deposition.\\The FESEM cross-sectional image shown in Figure \ref{fig:thickness} indicates that the film thickness lies in the nanometer range. A marginal increase in thickness from 181.2$\pm$0.6 nm to 189.1$\pm$0.5 nm is observed as the Ni doping concentration increases from 1\% to 4\%. This increase may be attributed to the higher deposition rate and the incorporation of Ni ions into the $\mathrm{Cd_{1-x}Mn_xS}$ lattice, which can promote additional material build-up during film growth. It may also be related to changes in nucleation and growth kinetics caused by Ni doping.
\subsection{Optical analysis}
\begin{figure}[ht]
	\centering
	\includegraphics[width=0.8\textwidth,height=0.4\textheight]{"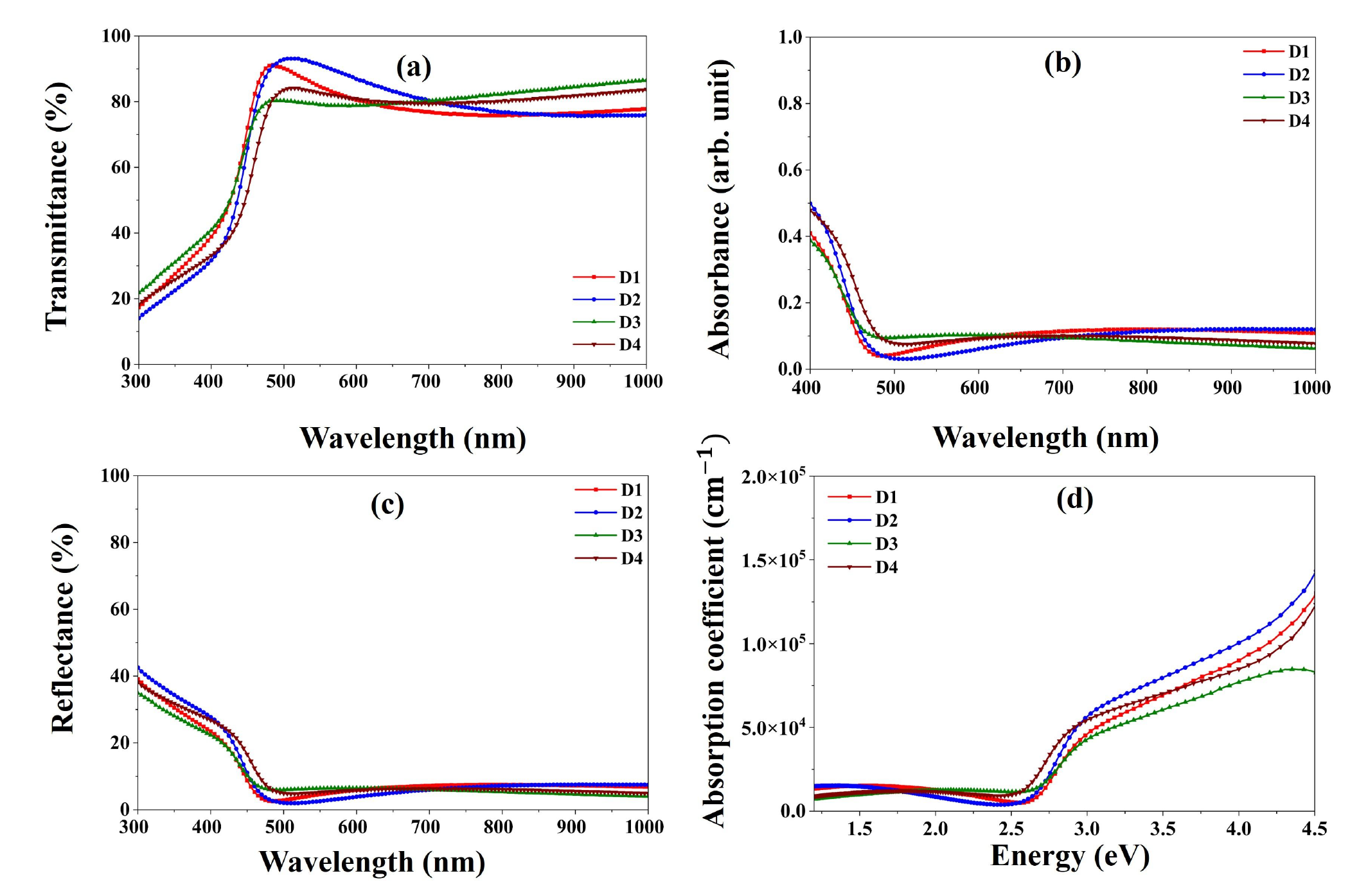"}
	\caption{(a) Transmittance versus wavelength, (b) Absorbance versus wavelength, (c) Reflectance versus wavelength, and (d) Absorption coefficient versus energy plot of Ni doped $\mathrm{Cd_{1-x}Mn_xS}$ thin films}
	\label{fig:absorbtion_transmittance}
\end{figure}

\begin{figure}[ht]
	\centering
	\includegraphics[width=0.8\textwidth,height=0.4\textheight]{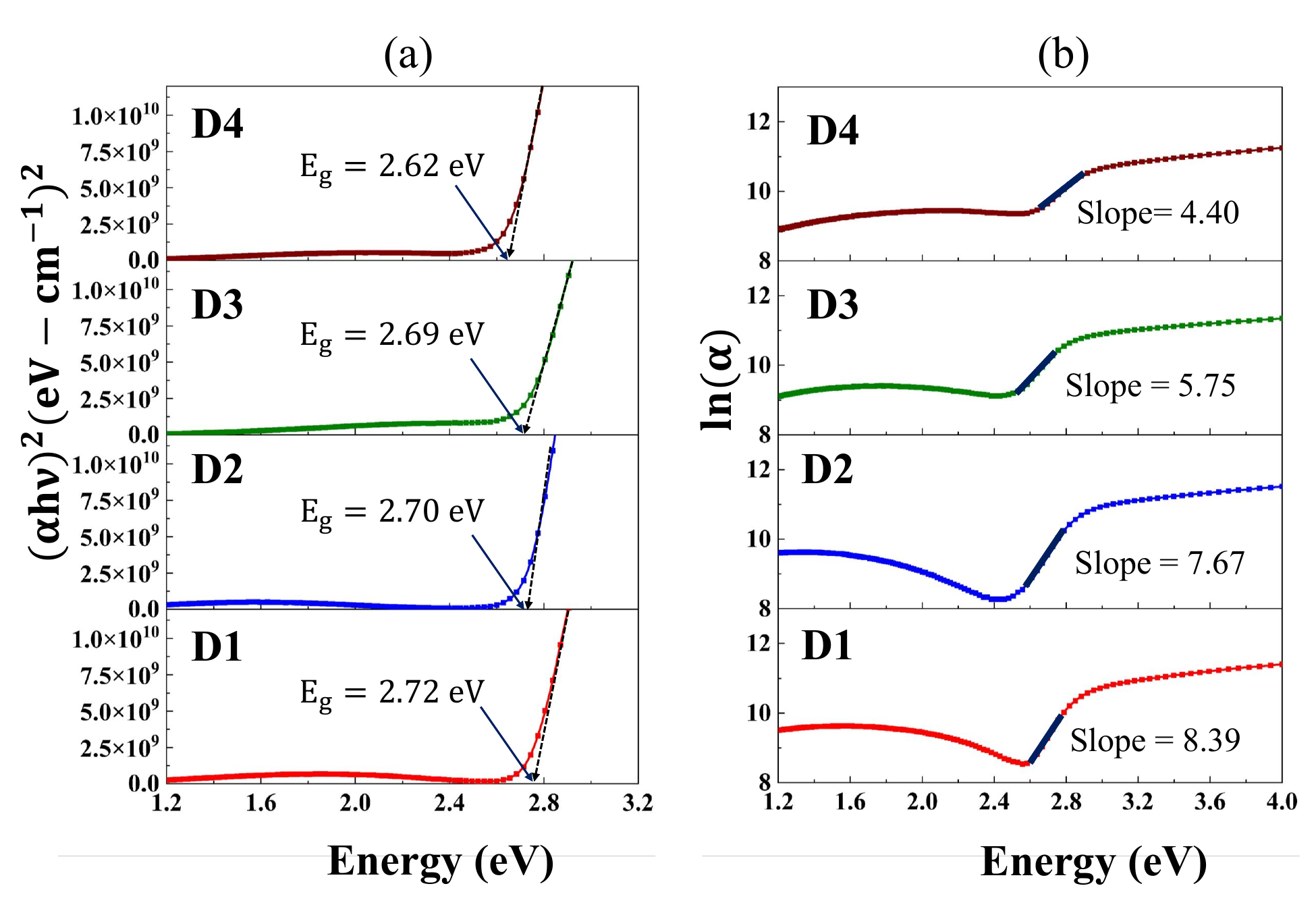}
	\caption{(a) $\mathrm{\left(\upalpha h \upnu\right)^2}$ versus energy and (b) $\ln(\upalpha)$ versus energy plot of Ni doped $\mathrm{Cd_{1-x}Mn_xS}$ thin films}
	\label{fig:band_gap_urbach}
\end{figure}

\begin{figure}[ht]
	\centering
	\includegraphics[width=0.8\textwidth,height=0.2\textheight]{"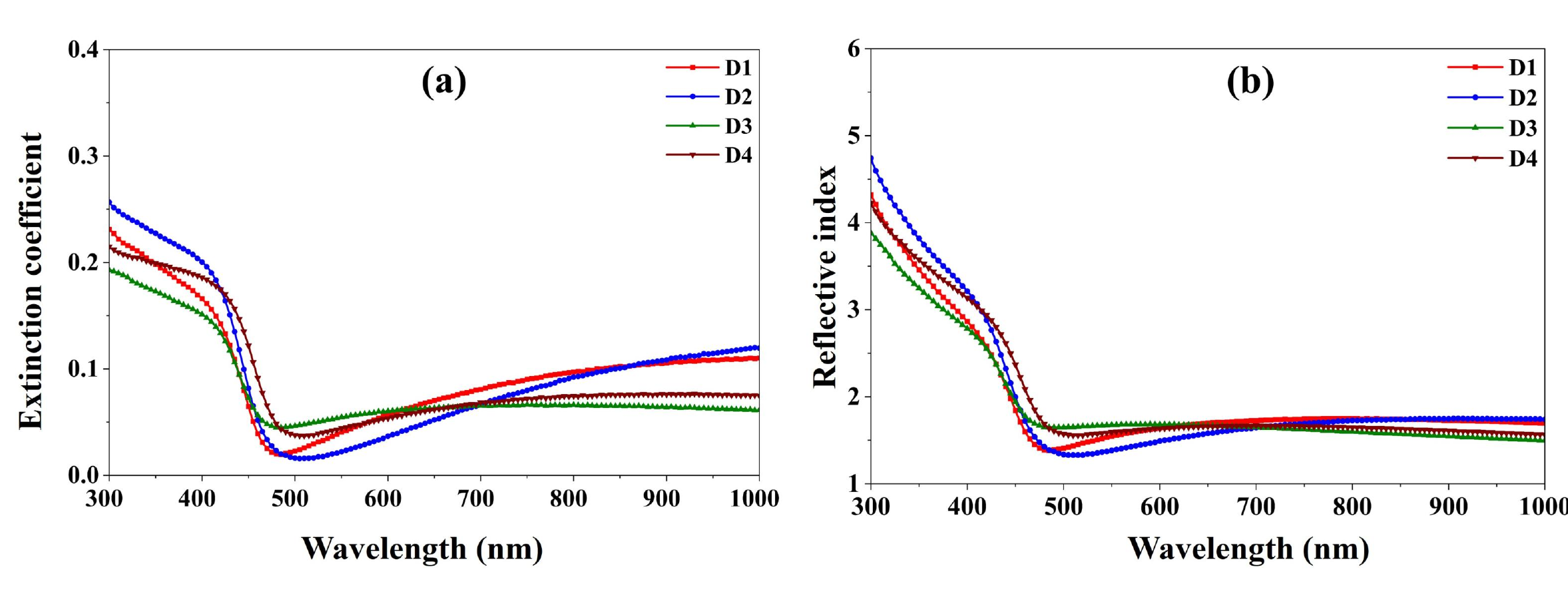"}
	\caption{(a) Extinction coefficient versus wavelength and (b) Refractive index versus wavelength of Ni doped $\mathrm{Cd_{1-x}Mn_xS}$ thin films}
	\label{fig:Extinction_refractive}
\end{figure}

\begin{figure}[ht]
	\centering
	\includegraphics[width=0.8\textwidth,height=0.4\textheight]{"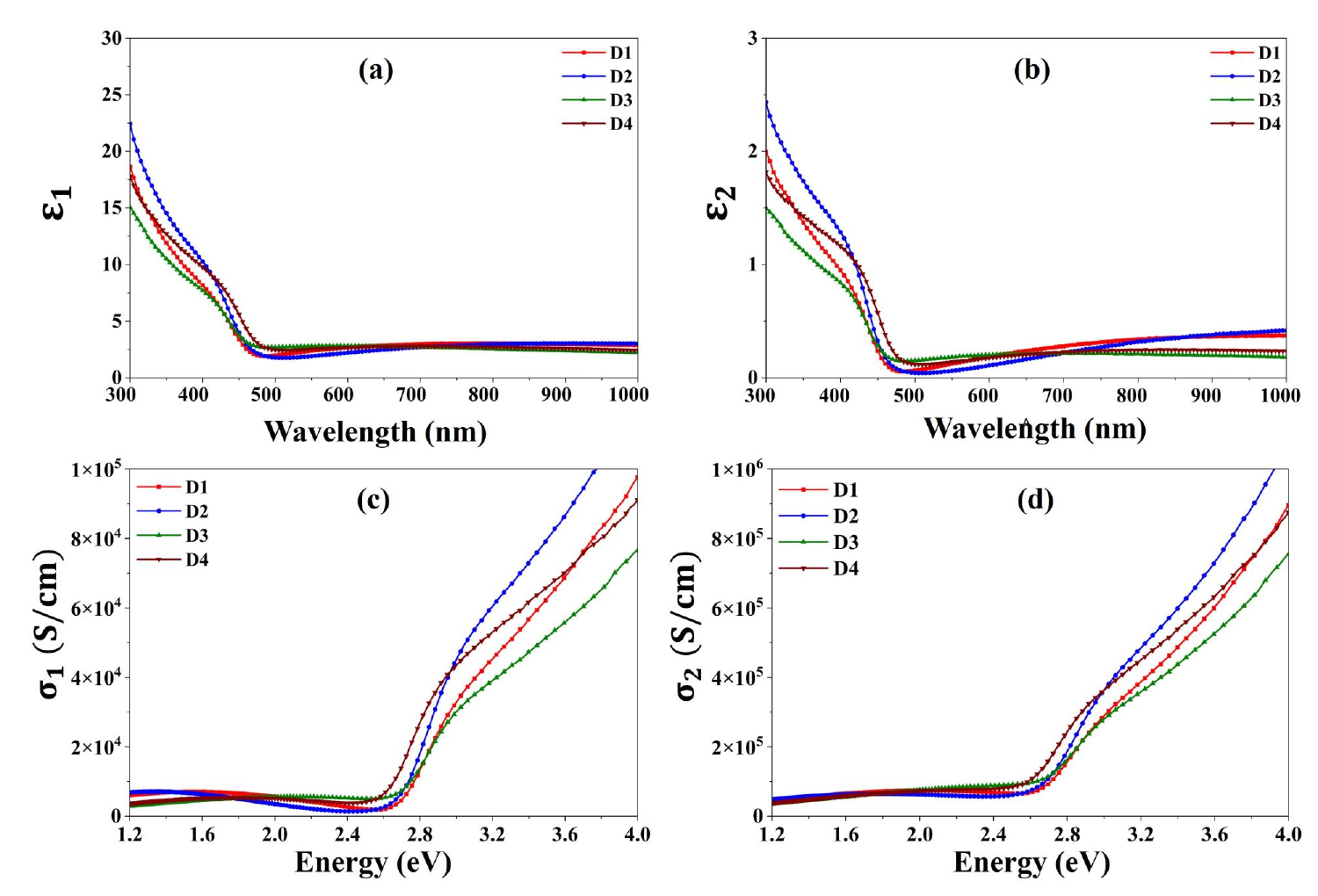"}
	\caption{(a) (a) Real part of the dielectric constant and (b) imaginary part of the dielectric constant as a function of wavelength; (c) real part of the optical conductivity and (d) imaginary part of the optical conductivity as a function of photon energy for Ni doped $\mathrm{Cd_{1-x}Mn_xS}$ thin films}
	\label{fig:Real_imaginary_dielectric}
\end{figure}

\begin{figure}[ht]
	\centering
	\includegraphics[width=0.8\textwidth,height=0.4\textheight]{"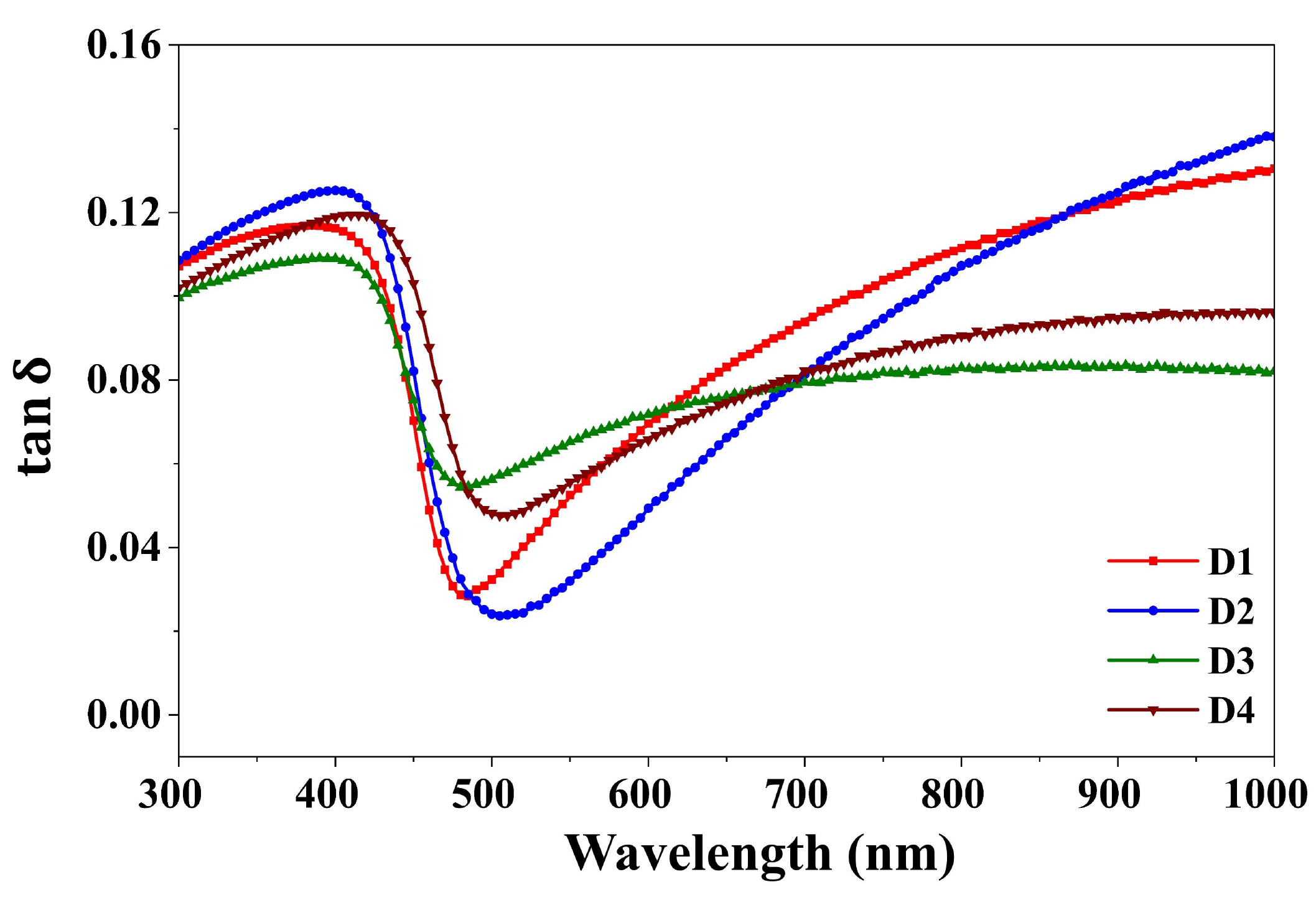"}
	\caption{$\tan\updelta$ versus wavelength plot for Ni doped $\mathrm{Cd_{1-x}Mn_xS}$ thin films}
	\label{fig:dissipation}
\end{figure}

Figure \ref{fig:absorbtion_transmittance} (a) shows the variation of transmittance with wavelength for the Ni-doped thin films. The films exhibit high transmittance, in the range of approximately 75–90\%, across wavelengths from nearly 480 nm to 1000 nm, covering the visible and near-infrared regions. This high transparency indicates that the films are suitable as window layer materials for thin-film solar cells (TFSCs). \\The absorbance (A) of the thin films was calculated using Eq. \ref{eq:absorbance} \cite{Das}.
\begin{equation}
	\mathrm{A=\log_{10}{\left(\frac{1}{T}\right)}}
	\label{eq:absorbance}
\end{equation}
where T is the transmittance value of the thin film.
Figure \ref{fig:absorbtion_transmittance} (b) presents the absorbance as a function of wavelength, showing that the absorbance remains low across wavelengths from nearly 480 nm to 1000 nm. Neglecting scattering losses, the reflectance of the thin films was calculated using Eq. \ref{eq:reflectance}, and the corresponding plot in Figure \ref{fig:absorbtion_transmittance} (c) confirms that the films possess low reflectance in the wavelength range of nearly 480 nm to 1000 nm \cite{boruah2025structural}.
\begin{equation}
	\mathrm{T=(1-R)^2\exp(-A)}
	\label{eq:reflectance}
\end{equation} 
Furthermore, the transmission edge shifts toward higher wavelengths with increasing Ni doping concentration, indicating a reduction in the optical band gap of the thin films. Similar behavior has been reported by Noor Shahina Begum et al. for Ni-doped $\mathrm{TiO_2}$ thin films \cite{begum2008effects} and  by Chao-Ming Huang et al. for Ni-doped ZnS thin films \cite{huang2009effect}.\\Using Eq. \ref{eq:absorption_coefficient}, the absorption coefficient of the thin films was calculated \cite{gogoi2021investigation}. Figure \ref{fig:absorbtion_transmittance} (d) shows the variation of the absorption coefficient as a function of photon energy. It is evident from the plot that the absorption coefficient lies in the range of $\mathrm{10^4-10^5 \,cm^{-1}}$, indicating the direct band gap nature of the thin films. 
\begin{equation}
	\upalpha=2.303\mathrm{\frac{A}{t}}
	\label{eq:absorption_coefficient}
\end{equation}
where A and t are the absorbance and thickness of the thin film respectively.
In many semiconducting materials, the optical absorption near the fundamental edge is commonly analyzed using band-to-band transition theory . Within this framework, the absorption data follow the Tauc power-law relation , given by Eq. \ref{eq:tauc} \cite{Sahin, Aslan}:
\begin{equation}
	\mathrm{\left(\upalpha h \upnu\right)^\frac{1}{n}=A(h\upnu-E_g)}
	\label{eq:tauc}
\end{equation}
where $\upalpha$ is the absorption coefficient, h$\upnu$ is the photon energy, A is a material-dependent constant and $\mathrm{E_g}$ is the energy band gap of the material. The parameter n depends on the nature of the electronic transition and takes different values, such as: 1/2 for direct allowed transition, 2/3 for direct forbidden transition, n=2 for indirect allowed transitions and n=3 for indirect forbidden transitions \cite{Arandhara,gogoi2021influence}. The Tauc plots corresponding to various electronic transition mechanisms were systematically analyzed to determine the nature of the optical transition. Among the different models, the direct allowed transition exhibited the best linearity in the $(\upalpha\mathrm{h}\upnu)^2$ versus h$\upnu$ plot. The optical band gap ($\mathrm{E_g}$) was estimated by extrapolating the linear region of this plot to the photon energy axis [$(\upalpha\mathrm{h}\upnu)^2=0]$ as shown in Figure \ref{fig:band_gap_urbach}. The reliability and accuracy of the fitting procedure are confirmed by the high coefficient of determination ($\mathrm{R^2}$) values, which lie in the range of 0.998–0.999, indicating an excellent agreement with the direct band gap model. The optical band gap of the $\mathrm{Cd_{1-x}Mn_xS}$ thin films decreases with increasing Ni doping concentration. The optical band gap of the Ni-doped $\mathrm{Cd_{1-x}Mn_xS}$ thin films is found to be reduced compared to that of the undoped $\mathrm{Cd_{1-x}Mn_xS}$ thin films ($\mathrm{E_g}=2.89$ eV), as previously reported \cite{pathok2025effect}. The corresponding band gap values are listed in Table \ref{tab:band_gap_urbach_energy}. The observed reduction in the optical band gap with increasing Ni concentration may be attributed to the incorporation of Ni ions into the $\mathrm{Cd_{1-x}Mn_xS}$ host lattice, which can induce lattice distortion, introduce localized defect states, and modify the electronic structure of the material. In general, Ni centres in wide band-gap II–VI semiconductors such as CdS, ZnS, and ZnO are known to possess shallow states due to their amphoteric nature, giving rise to both donor- and acceptor-type levels. In particular, $\mathrm{Ni^{2+}}$ ions occupying cationic sites in CdS have been reported to form deep acceptor states above the valence band \cite{Heitz}. The presence of such impurity states within the forbidden gap facilitates lower-energy optical transitions, thereby contributing to the observed band gap narrowing in the present Ni-doped $\mathrm{Cd_{1-x}Mn_xS}$ thin films. A similar decrease in band gap energy with increasing Ni concentration has also been reported by A. Rmili et al. and S. Chandramohan for Ni-doped CdS thin films \cite{rmili2013structural, Chandramohan}.\\The Urbach energy of the thin films was evaluated using the empirical Urbach relation given in Eq. \ref{eq:urbach}. 
\begin{equation}
	\ln{\upalpha}=\ln{\upalpha_0}+\frac{\mathrm{h}\upnu}{\mathrm{E_u}}
	\label{eq:urbach}
\end{equation}
where $\upalpha$ denotes absorption coefficient, $\upalpha_0$ represents the pre-exponential factor, h$\upnu$ represents the energy and $\mathrm{E_u}$ represents the urbach energy of the thin films.
The Urbach energy was determined from the inverse slope of the linear region of the $\ln(\upalpha)$ versus photon energy (h$\upnu$) plot, as depicted in Figure \ref{fig:band_gap_urbach}. The extracted Urbach energy values are summarized in Table \ref{tab:band_gap_urbach_energy}. An increase in Urbach energy with increasing Ni doping concentration is observed, indicating the broadening of band-tail states and an increase in the density of localized states near the band edges in the $\mathrm{Cd_{1-x}Mn_xS}$ thin films. Urbach tails are formed due to electron–phonon and electron–defect interactions, and the Urbach parameter is strongly dependent on temperature and doping concentration \cite{Sarangapani, Jain}. Although Ni incorporation promotes crystal growth and enhances the degree of crystallinity, substitutional Ni ions introduce compositional fluctuations and impurity-induced electronic disorder, giving rise to localized states extending into the forbidden energy gap. The broadening of these band-tail states facilitates optical transitions involving lower photon energies, thereby accounting for the observed reduction in the optical band gap with increasing Ni content.
\begin{table}[ht]
	\centering
	\renewcommand{\arraystretch}{1.2}
	\setlength{\tabcolsep}{24pt}
	\begin{threeparttable}
		\caption{Estimated band gap and urbach energy of the Ni doped thin films}
		\label{tab:band_gap_urbach_energy}
		\begin{tabularx}{\linewidth}{c c c c}
			\hline
			\text{Sample code} & \makecell{Band gap energy\\(eV)} & \makecell{Urbach energy\\(eV)} & \makecell{Thickness\\($\mathrm{nm}$)}\\
			\hline
			D1 & 2.72 & 0.12 & 181.2$\pm$0.8\\
			D2 & 2.70 & 0.13 & 182.6$\pm$1.2\\
			D3 & 2.69 & 0.17 & 188.3$\pm$0.6\\
			D4 & 2.62 & 0.23 & 189.1$\pm$0.7\\
			\hline
		\end{tabularx}
		\begin{tablenotes}
			\footnotesize
			\item \textit{The reported thickness values represent the mean $\pm$ standard error, obtained from ten independent measurements performed across 800 nm range of the thin film.}
		\end{tablenotes}
	\end{threeparttable}
\end{table}
Figure \ref{fig:Extinction_refractive} shows the variation of the absorption index, or extinction coefficient (k), calculated using Eq. \ref{eq:extinction} \cite{gode2011annealing}, as a function of wavelength ($\uplambda$). The extinction coefficient is very low in the visible and near-infrared (NIR) regions, which is consistent with the high transmittance observed for the thin films.
\begin{equation}
	\mathrm{k=\frac{\upalpha\uplambda}{4\uppi}}
	\label{eq:extinction}
\end{equation}
where $\upalpha$ is the absorption coefficient of the thin film.\\Eq. \ref{eq:refractive_index} was used to calculate the refractive index (n) of the thin films \cite{kariper2011structural}, and the corresponding variation with wavelength is shown in Figure \ref{fig:Extinction_refractive}. The refractive index lies in the range of 1.5 to 2.3 over the wavelength interval nearly 480–1000 nm, corresponding to the visible and near-infrared (NIR) regions. Below the absorption-edge wavelength, the refractive index exhibits anomalous dispersion, which may be attributed to the strong absorption near the band edge and the associated rapid variation in electronic polarization.
\begin{equation}
	\mathrm{n=\frac{1-R}{1+R}+\sqrt{\frac{4R}{(1-R)^2}-k^2}}
	\label{eq:refractive_index}
\end{equation}
where R is the reflectance and k is the absorption index of the thin film.
The interaction of electromagnetic radiation with a material is fundamentally governed by its complex dielectric constant ($\upvarepsilon^*(\upomega)=\upvarepsilon_1(\upomega)+\mathrm{i}\upvarepsilon_2(\upomega)$) and complex optical conductivity ($\upsigma^*(\upomega)=\upsigma_1(\upomega)+\mathrm{i}\upsigma_2(\upomega)$), both of which exhibit strong angular frequency ($\upomega$) dependence \cite{aly2023adjusting}. The real part of the optical conductivity ($\upsigma_1$) is directly linked to the imaginary component of the dielectric constant ($\upvarepsilon_2$) and characterizes the dissipative processes associated with the absorption of electromagnetic radiation. In contrast, the imaginary part of the optical conductivity ($\upsigma_2$) is related to the real part of the dielectric constant ($\upvarepsilon_1$) and reflects the dispersive response arising from polarization mechanisms within the material.\\These interrelated parameters provide a comprehensive description of the optical response, encompassing both energy dissipation and storage processes. Furthermore, they serve as critical indicators of the underlying electronic structure and charge carrier dynamics, thereby offering valuable insight into the dispersive behavior and optoelectronic performance of the material. 
The real ($\upvarepsilon_1$) and imaginary ($\upvarepsilon_2$) components of the complex dielectric function were evaluated using Eqs. \ref{eq:real_dielectric} and \ref{eq:imaginary_dielectric}, respectively \cite{ulutas2020effects}. 
\begin{equation}
	\upvarepsilon_1=\mathrm{n^2-k^2}
	\label{eq:real_dielectric}
\end{equation}
\begin{equation}
	\upvarepsilon_2=\mathrm{2nk}
	\label{eq:imaginary_dielectric}
\end{equation}
where n and k are the refractive index and absorption index of the thin films.\\Figures \ref{fig:Real_imaginary_dielectric} (a) and \ref{fig:Real_imaginary_dielectric} (b) illustrate the wavelength dependence of $\upvarepsilon_1$ and $\upvarepsilon_2$ for the Ni-doped $\mathrm{Cd_{1-x}Mn_xS}$ thin films. It is observed that ($\upvarepsilon_1$) remains significantly higher than ($\upvarepsilon_2$) throughout the investigated spectral range, indicating that the polarization response of the films dominates over absorptive losses. The comparatively low magnitude of ($\upvarepsilon_2$) suggests weak optical absorption and reduced dielectric loss, implying minimal energy dissipation within the films. Consistent with this behavior, the $\upsigma_1$ is lower than the $\upsigma_2$ as shown in Figures \ref{fig:Real_imaginary_dielectric} (c) and \ref{fig:Real_imaginary_dielectric} (d), which indicates that the films store more electromagnetic energy through polarization than they dissipate through absorption.
The dissipation factor, $\tan\updelta$, is the ratio of power lost in a dielectric material to the power transmitted through it, and it reflects the imperfection of the dielectric. Most dielectric materials have a low dissipation factor, which is desirable because it reduces energy loss as heat. The dissipation factor can be calculated using Eq. \ref{eq:dissipation} \cite{Fuchs}:
\begin{equation}
	\tan\updelta=\frac{\upvarepsilon_2}{\upvarepsilon_1}
	\label{eq:dissipation}
\end{equation} Figure \ref{fig:dissipation} shows the variation of the dissipation factor with wavelength for the thin films. It is observed that the dissipation factor is small throughout the measured range, indicating low dielectric losses and minimal energy dissipation in the films. A pronounced minimum in the dissipation factor is observed in the 480–510 nm range for all thin films. This behavior suggests reduced dielectric loss near the absorption edge, where optical transitions become less significant and the loss tangent $\tan\updelta$ reaches a minimum. In this region, the real part of the dielectric function remains dominant over the imaginary part, indicating that polarization effects are stronger than absorptive losses \cite{ALHARBI2023107984}.
\subsubsection{Hall measurement and Electrical analysis}
Employing the Van der Pauw Hall effect configuration under the constant applied magnetic field of 2500 G and a constant current of 60 nA at room temperature, the carrier concentration, mobility, Hall coefficient and sheet resistance of the Ni-doped thin films were determined \cite{Ohring, Chopra, Cieśliński}. The calculated Hall parameters are summarized in Table \ref{tab:Hall_effect}. As the Ni doping concentration increases from 1 to 4\%, the sheet resistance decreases from 4.68 $\times 10^8$ to 3.67 $\times 10^8\,\Omega/\square$, whereas both the carrier concentration and carrier mobility exhibit an increasing trend. This may be attributed to the improved crystallinity and increased grain size, which leads to a reduction in grain boundaries \cite{Addonizio,Mustajab}. A similar pattern was also observed for Ni-doped CdS thin films \cite{Premarani}.  The negative values of the Hall coefficient obtained for all the samples confirm the n-type conductivity of the thin films \cite{Pathok}. Moreover, the polarity of the charge carriers remains unaffected over the investigated Ni doping range.

\begin{table}[ht]
	\centering
	\renewcommand{\arraystretch}{1.2}
	\setlength{\tabcolsep}{11pt}
		\caption{Estimated Hall parameters of Ni doped $\mathrm{Cd_{1-x}Mn_xS}$ thin films}
		\label{tab:Hall_effect}
		\begin{tabularx}{\linewidth}{c c c c c}
			\hline
			\text{Sample code} & \makecell{Sheet resistance\\($\times10^8\,\Omega/\square$)} & \makecell{Carrier concentration\\($\times10^{13}$ cm$^{-3}$)} & \makecell{Mobility, $\upmu$\\($\mathrm{cm^2/V.s}$)} & \makecell{Hall coefficient\\ ($\times10^4~\mathrm{cm^3/C}$)}\\
			\hline
			D1 & 4.68 & 7.16 & 46.71 & -8.73\\
			D2 & 4.29 & 7.47 & 48.34 & -8.36\\
			D3 & 3.85 & 7.99 & 48.88 & -7.82\\
			D4 & 3.67 & 8.21 & 49.77 & -7.61\\
			\hline
		\end{tabularx}
	\end{table}

\begin{figure}[ht]
	\centering
	\includegraphics[width=0.8\textwidth,height=0.4\textheight]{"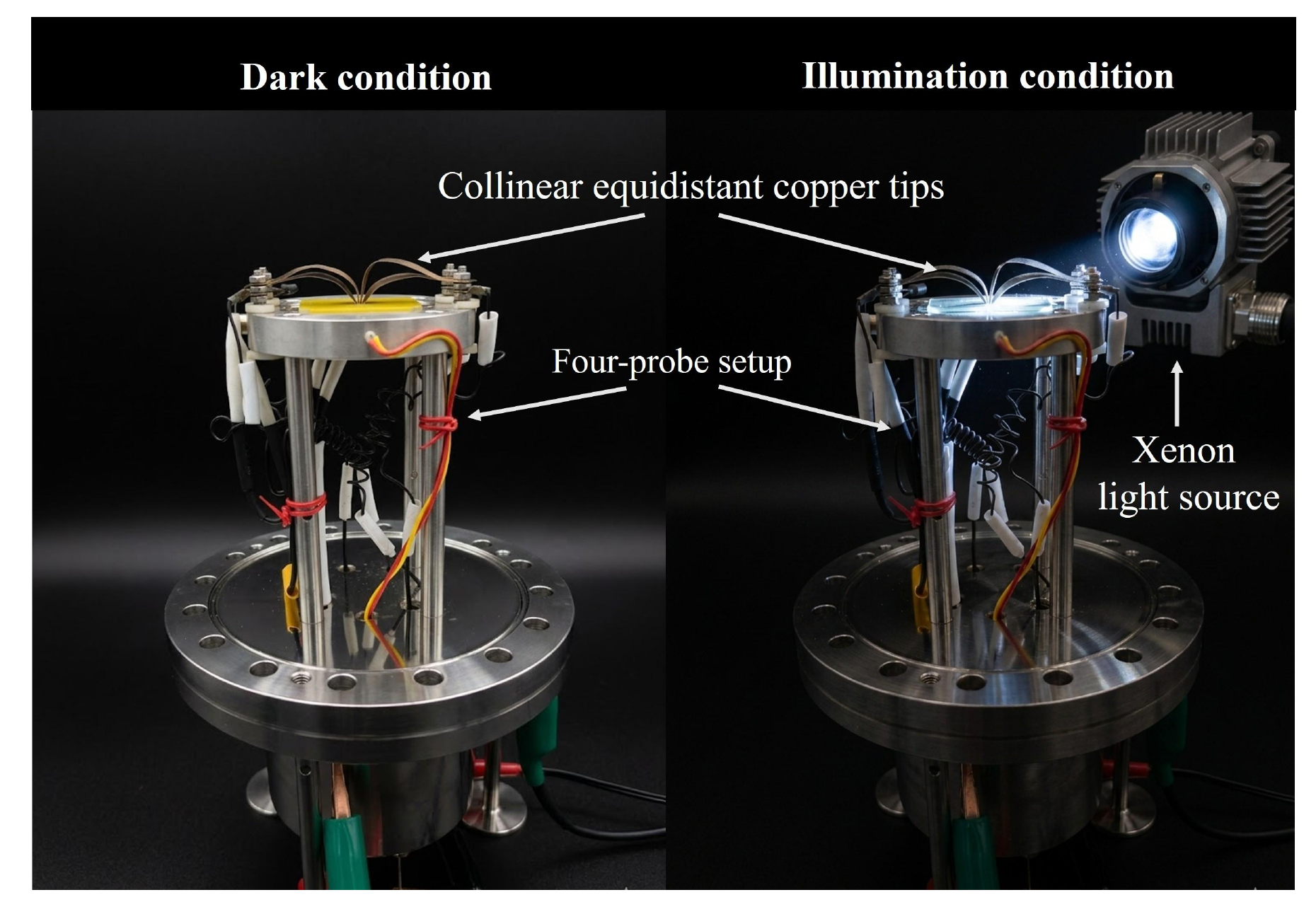"}
	\caption{Four-probe setup for I-V characteristics measurement of the thin films}
	\label{fig:I_V_set_up}
\end{figure}

The electrical parameters of the thin films were evaluated using a four-probe setup in conjunction with a source meter. The collinear probes were separated by an interprobe distance of 2 mm. During the measurement, a xenon lamp with an intensity of 15 mW$\mathrm{cm^{-2}}$ was employed for illumination as shown in Figure \ref{fig:I_V_set_up}. 

\begin{figure}[ht]
	\centering
	\includegraphics[width=0.8\textwidth,height=0.4\textheight]{"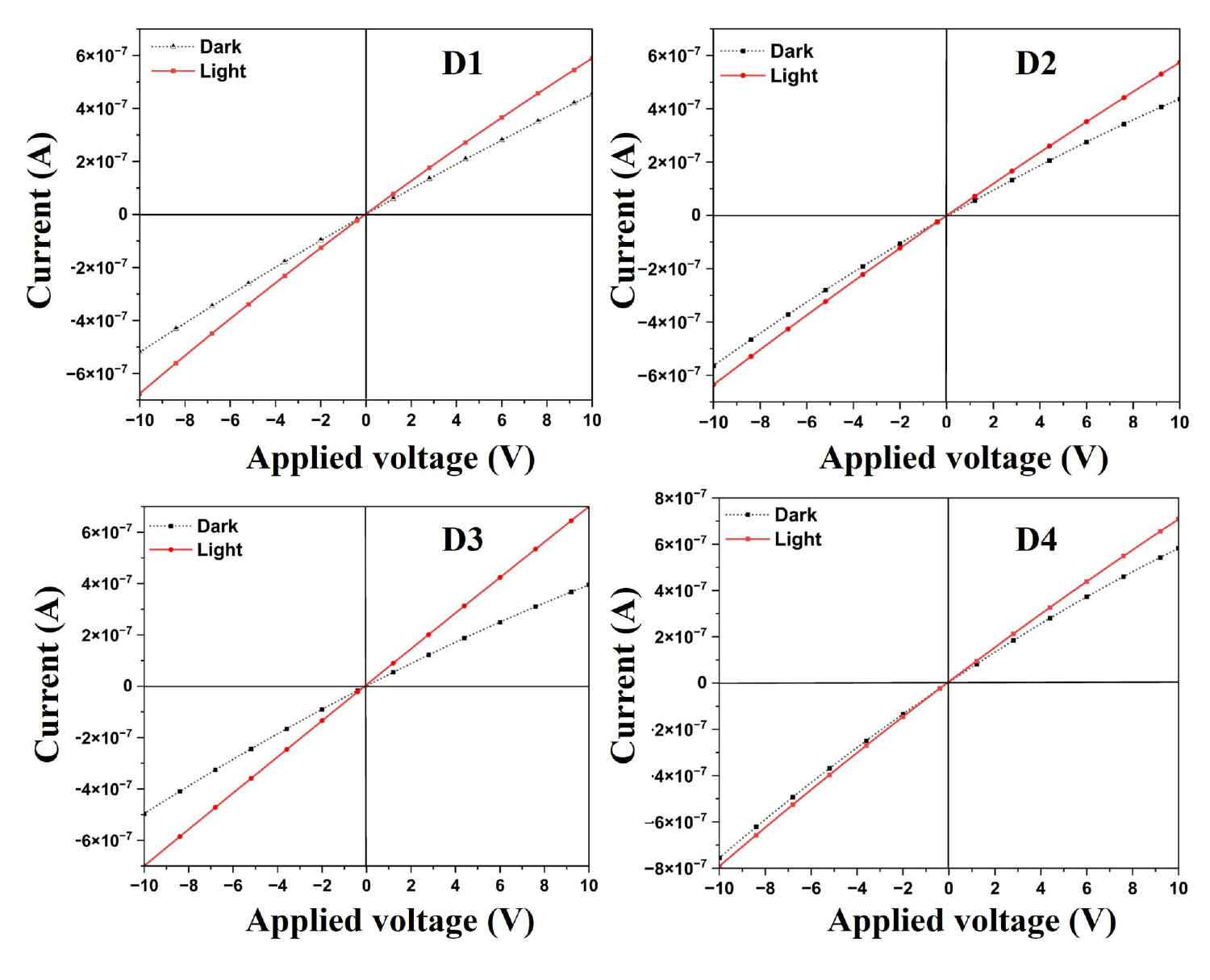"}
	\caption{I-V characteristics of Ni doped $\mathrm{Cd_{1-x}Mn_xS}$ thin films}
	\label{fig:I-V}
\end{figure}

The current–voltage (I–V) characteristics of the Ni-doped $\mathrm{Cd_{1-x}Mn_xS}$ thin films under dark and illuminated conditions are presented in Figure \ref{fig:I-V}. The linear nature of the plot indicates ohmic behavior, confirming the absence of a Schottky barrier at the interface between the semiconducting thin film and the metal electrode contacts. The measurements were carried out at room temperature and atmospheric pressure (NTP). The resistance values obtained from the plot are listed in Table \ref{tab:I-V-data}. Since the substrate used for thin-film deposition is non-conducting glass, the resistivity of the deposited films ($\uprho$) can be determined using Eq. \ref{eq:resistivity} \cite{ali2018enhancement}:
\begin{equation}
	\uprho=\frac{\uppi}{\ln2}\mathrm{\frac{V}{I}{t}}
	\label{eq:resistivity}
\end{equation}
where t represents the thickness of the thin film. The resistivity values of the thin films under both dark and illuminated conditions are summarized in Table \ref{tab:I-V-data}. A systematic decrease in resistivity with increasing Ni doping concentration is observed, which can be attributed to the enhanced charge carrier concentration and improved charge transport induced by Ni incorporation. Furthermore, the resistivity under illumination is consistently lower than that in the dark, indicating the generation of additional photocarriers upon light exposure.

\begin{figure}[ht]
	\centering
	\includegraphics[width=0.8\textwidth,height=0.4\textheight]{"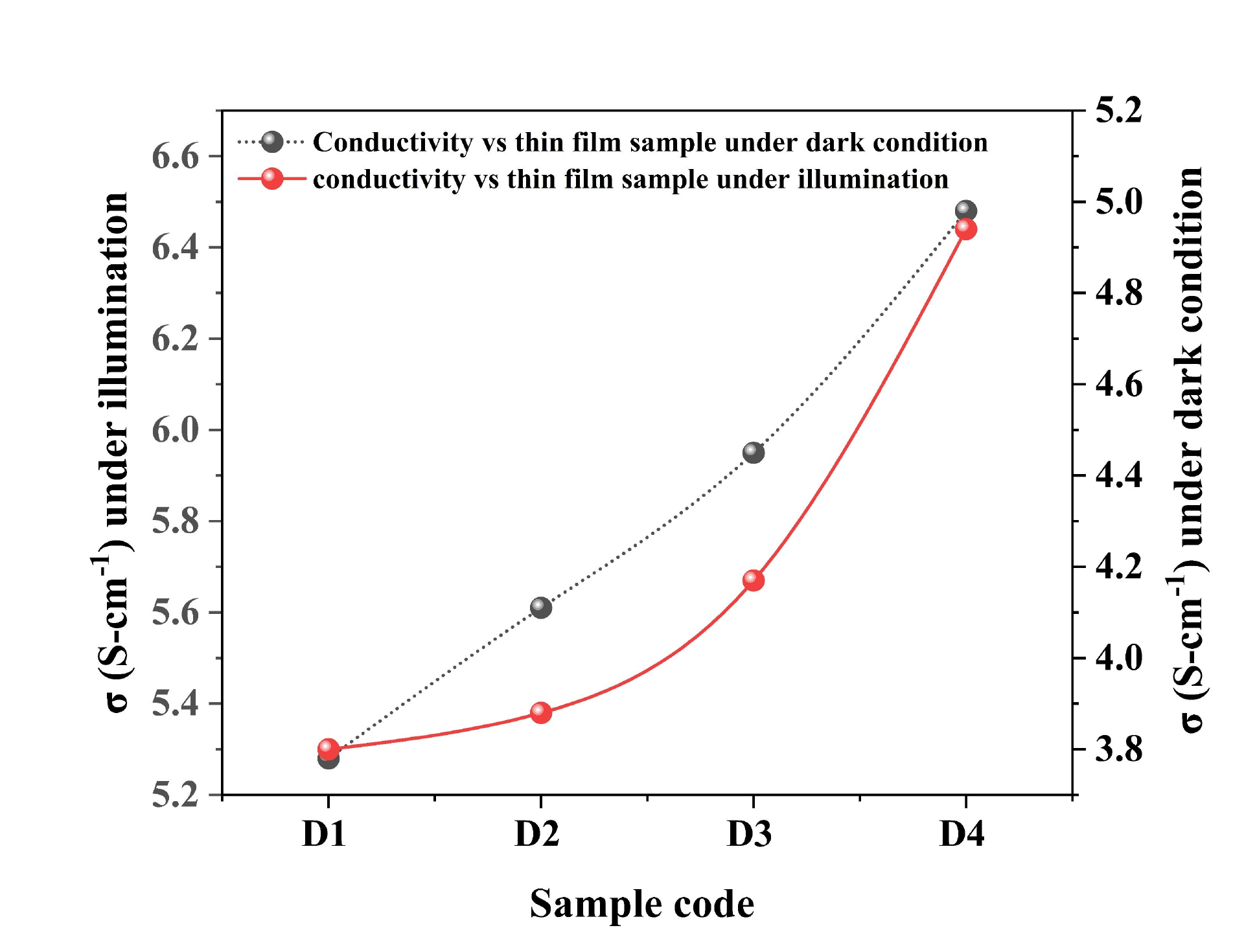"}
	\caption{Conductivity of Illuminated and dark condition versus sample code of Ni doping  $\mathrm{Cd_{1-x}Mn_xS}$ thin films}
	\label{fig:I-V_dark_illuminated}
\end{figure}

Correspondingly, the electrical conductivity($\upsigma=\frac{1}{\uprho}$) of the thin films increases with increasing Ni doping concentration. The illuminated films exhibit significantly higher conductivity compared to those measured under dark conditions, further confirming the photoconductive nature of the material. As shown in Figure \ref{fig:I-V_dark_illuminated}, both dark and photo-induced electrical conductivities increase monotonically with increasing Ni content, highlighting the role of Ni doping in enhancing carrier generation and transport.
\begin{table}[ht]
	\centering
	\renewcommand{\arraystretch}{1.8}
	\setlength{\tabcolsep}{3.3 pt}
	\caption{Estimated structural parameters of Ni doped $\mathrm{Cd_{1-x}Mn_xS}$ thin films}
	\label{tab:I-V-data}
	\begin{tabularx}{\linewidth}{c c c c c c c }
		\hline
		\makecell{\text{Sample}\\\text{Code}} & 
		\multicolumn{3}{c}{\text{Illuminated condition}} & \multicolumn{3}{c}{\text{Dark condition}}\\
		\cline{2-7}
		& \makecell{\text{Resistance}\\($\times 10^7\,\upOmega$)} & \makecell{\text{Resistivity}\\$(\times 10^3\,\upOmega\mathrm{-cm})$} & \makecell{\text{Conductivity}\\$\left(\times 10^{-4} \,\mathrm{S-cm^{-1}}\right)$} &\makecell{\text{Resistance}\\($\times 10^{7}\,\upOmega$)} & \makecell{\text{Resistivity}\\$(\times 10^3\,\upOmega\mathrm{-cm})$} & \makecell{\text{Conductivity}\\$\left(\times 10^{-4} \,\mathrm{S-cm^{-1}}\right)$} \\
		\hline
		D1 & 1.71 & 1.89 & 5.28 & 2.45 & 2.63 & 3.80 \\
		D2 & 1.36 & 1.78 & 5.61 & 2.24 & 2.58 & 3.88 \\
		D3 & 1.39 & 1.68 & 5.95 & 2.16 & 2.40 & 4.17 \\
		D4 & 1.44 & 1.54 & 6.48 & 1.68 & 2.03 & 4.94 \\
		\hline
	\end{tabularx}
\end{table} 
For comparison, the electrical conductivity of the Ni-doped $\mathrm{Cd_{1-x}Mn_xS}$ thin films was evaluated alongside that of several commonly used window-layer materials. As summarized in the Table \ref{tab:electrical conductivity_window_layer}, the Ni-doped $\mathrm{Cd_{1-x}Mn_xS}$ films exhibit significantly higher electrical conductivity than the undoped $\mathrm{Cd_{1-x}Mn_xS}$ films and many other window layer materials. This enhancement can be attributed to the increased charge carrier concentration and improved charge transport arising from Ni incorporation.
\begin{table}[H]
	\centering
	\renewcommand{\arraystretch}{1.2}
	\setlength{\tabcolsep}{8pt}
	\caption{Comparison of electrical conductivity of different window layer materials}
	\label{tab:electrical conductivity_window_layer}
	\begin{tabularx}{\linewidth}{c c c c c}
		\hline
		Sr. no. & \makecell{Thin film} & \makecell{Electrical conductivity\\($\mathrm{S-cm^{-1}}$)} & Deposition method & \makecell{Reference} \\
		\hline
		1 & CdS &  8.33$\times 10^{-4}$ & Vacuum evaporation & \cite{al1998influence} \\
		2 & CdS & 2.28$\times 10^{-6}$ & Chemical bath deposition & \cite{kariper2011structural}\\
		3 & MnS & 1.60 $\times 10^{-7}$ & Spray pyrolysis & \cite{ibrahim2021physical}\\
		4 & MnS & 1.41$\times 10^{-6}$ & Chemical bath deposition & \cite{ezema2007analysis}\\
		5 & CdSSe & 1.60$\times 10^{-5}$ & Chemical bath deposition & \cite{das2024insight}\\
		6 & ZnS & $10^{-9}$ & Chemical  bath deposition & \cite{gode2011annealing}\\
		7 & ZnSSe & 2.27$\times 10^{-4}$ & Chemical  bath deposition & \cite{boruah2025low}\\
		8 & CdZnS & 4.31$\times 10^{-4}$ & Chemical  bath deposition & \cite{borah2025effect}\\
		9 & $\mathrm{Cd_{1-x}Mn_xS}$ & 3.43$\times 10^{-4}$ & Chemical bath deposition & \cite{pathok2025effect}\\
		10 & $\mathrm{Ni:Cd_{1-x}Mn_xS}$ & 4.94$\times 10^{-4}$ & Chemical bath deposition & present work\\
		\hline
	\end{tabularx}
\end{table} 
\section{Conclusion}
Ni-doped $\mathrm{Cd_{1-x}Mn_xS}$ thin films were successfully deposited via the chemical bath deposition technique, and the influence of Ni incorporation on their structural, morphological, optical, and electrical properties was systematically examined. X-ray diffraction and high-resolution transmission electron microscopy analyses confirmed the polycrystalline nature of the films with a cubic zinc blende structure. A gradual increase in crystallite size accompanied by a reduction in microstrain and dislocation density with increasing Ni concentration indicates an overall improvement in crystalline quality. The successful incorporation of Ni into the $\mathrm{Cd_{1-x}Mn_xS}$ lattice was verified through compositional analysis, while FESEM analysis revealed dense, uniform, and well-adhered films with a slight increase in thickness upon doping. The average grain size of the spherical grains in the deposited thin films increases from 90.01$\pm$0.69 nm to 147.90$\pm$1.31 nm with increasing dopant concentration.  Optical studies demonstrated high transmittance in the visible and near-infrared regions (75-90\%), along with a systematic reduction in the optical band gap, suggesting band structure modification induced by Ni incorporation. Furthermore, the electrical conductivity of the films was found to increase with Ni doping and exhibited enhanced values under illumination, confirming the photoconductive nature of the material. The observed improvements in structural order, optical response, and charge transport characteristics highlight the significant role of Ni doping in tailoring the properties of $\mathrm{Cd_{1-x}Mn_xS}$ thin films.\\Consequently, the excellent structural, morphological, optical, and electrical characteristics of the Ni-doped $\mathrm{Cd_{1-x}Mn_xS}$ thin films suggest that they can be used in various optoelectronic applications.
\newpage
\noindent\textbf{CRedit author statement}: \\Himanshu Sharma Pathok: Conceptualization, methodology, validation, formal analysis, investigation, resources, and writing original draft. Padma Pani Shahu and Himanshu Kalita: Conceptualisation, writing- review and editing. Prasanta Kumar Saikia: Conceptualization, validation, writing-review and editing, supervision of the entire work.\\
\textbf{Acknowledgment}: \\The authors would like to acknowledge Dibrugarh University for providing the opportunity to carry out this research work. The authors are also grateful to SAIF, Gauhati University; IASST; and CSIR-NEIST, Jorhat, for their support in the analysis and characterization of the materials. The authors express their sincere thanks to Dr. A. T. T. Mostako (Dibrugarh University, Assan, India) for granting permission to use the I–V characteristics instrument and to Ms. Gargee Bhattacharyya (National Institute of Technology, Rourkela, Odisha, India), Ms. Vaishnabi Saikia (IIT Guwahati, Assam, India), Dr. Angona Kashyap (Dibru College, Dibrugarh, Assam, India), and Dr. Rulee Boruah (POWIET, Jorhat, Assam, India) for their valuable assistance in the synthesis and characterization processes.\\
\textbf{Funding}: \\No funding was received for this study.\\
\textbf{Data availability statement}: \\
The data supporting the findings of this study are available from the corresponding author upon reasonable request.\\
\textbf{Declaration of Interest}\\	The authors declare that they have no known competing financial interests or personal relationships that could have appeared to influence the work reported in this paper.
	
\end{document}